\documentclass[]{aa}  

\usepackage{graphicx}
\usepackage{subfigure}
\usepackage{txfonts}
\usepackage{tikz}
\usetikzlibrary{shapes.geometric,arrows}
\tikzstyle{startstop} = [rectangle, thick,rounded corners, minimum width=3cm, minimum height=1cm,text centered, draw=black, fill=red!30]
\tikzstyle{process} = [rectangle,thick, minimum width=3cm, minimum height=1cm, text centered, draw=black, fill=orange!30]
\tikzstyle{repeat} = [circle, thick, minimum width=1cm, minimum height=1cm, text centered, draw=black, fill=blue!30]
\tikzstyle{decision} = [rectangle, thick, minimum width=3cm, minimum height=1cm, text centered, draw=black, fill=green!30]
\tikzstyle{arrow} = [very thick,->,>=stealth]

\begin{document} 

  \title{Short term variability of DS Tuc A observed with TESS.}

   \author{S. Colombo
          \inst{1}
          \and
           A. Petralia \inst{1}
          \and
           G. Micela \inst{1}
          }

   \institute{INAF - Osservatorio Astronomico di Palermo, Piazza del Parlamento 1, Palermo, Italy.\\
              \email{salvatore.colombo@inaf.it}
             }

   \date{}

 
  \abstract
  {Impulsive short term variations occur in all kinds of solar-type stars.
  They are the results of complex phenomena such as the stellar magnetic field reconnection, low-level variability or in some cases even star-planet interactions. The radiation arising from these events is often highly energetic and, in stars hosting planets, may interact with the planetary atmospheres. Studying the rate of these energetic phenomena is fundamental to understand their role in modifying the chemical composition or, in some extreme cases, to the disruption of the planetary atmospheres.}
   {Here, we present a new procedure developed to identify the impulsive events in TESS light curves. Our goal is to have a  simple and effective tool to study the short-term activity of a star using only its light curve, in order to derive its distribution and energetic. As our first case, we studied the system DS Tuc.}
   {Our technique consists of fitting the TESS light curves using iteratively Gaussian processes in order to remove all the long-term stellar activity contributions. Then, we identify the impulsive events and, derive amplitudes, time scales and the amount of energy emitted.}
    {We validate our procedure using the AU Mic TESS light curves obtaining results consistent with those presented in the literature. We estimate the frequency distribution of energetic events for DS Tuc. In particular, we find that there are  $\approx 2$ events per day with energy greater than $2 \times 10^{32}$erg. 
    We find evidence for a favoured stellar phase for short term activity on AU Mic, and also indications of short term activity in phase with the planetary orbit.
    For DS Tuc we find that the events distribution is not equally spaced in time but often grouped. The resulting distribution may be used to estimate the impact of short term variability on planetary atmosphere chemical compositions.}
   {}


   \maketitle
%

\section{Introduction}
 Solar-type stars show activity on several time scales, including very short ones compared to planetary or stellar life timescales. The word "activity" involves several observed phenomena, including high energy (EUV-X-ray) variable radiation, that may be responsible for changes in chemistry and dynamics of planetary atmospheres.

 The study of the effects of stellar activity on planets is of great importance for understanding their role in defining planetary atmosphere properties and their chemical compositions.
 For example, flares generate a sensible increase of X and UV fluxes. This radiation affects the planetary atmosphere in different ways: UV and EUV radiations are responsible for molecular dissociation and photo-ionization in the upper layers, while X-rays induce secondary ionization affecting deeper layers \citep{Maloney1996, Cecchi2006, Locci2021}. The interaction between radiation and planetary atmosphere results in a change of the chemical composition, an increase in temperature and, in some extreme cases, atmospheric evaporation \citep{Watson1981, Erkaev2007}. These effects will change during the star life since stellar activity evolve significantly with the strongest effects at young ages.

 The frequency of energetic events, such as flares, is crucial for the chemical structure of the planetary atmosphere. Ionising radiation triggers changes in chemical composition with a timescale inversely proportional to the atmosphere density. If the time between two consecutive energetic events is smaller than the time needed from a specific species to go back to equilibrium, the emission produced by the subsequent event will trigger new chemistry in a highly non-linear path \citep{Venot2016,Chen2021}.
 In principle, to estimate the frequency of energetic events would require long - possibly contiguous - observations at EUV-X-ray wavelengths. These are very difficult to obtain but we can take advantage of the fact that, in general, short term events are observed in several bands and X-ray events are linked with short term variability in the optical band. For example, in a recent XMM-Newton observation DS Tuc A shows the occurrence of two twin X-ray flares detected also with the optical monitor (Pillitteri et al., in prep.). Therefore, as a rough assumption, we assume that  short term optical variability could be used as a tracer of short term variability at higher energy \citep{Flaccomio2018}.

 Moreover, finding periodicity in the short-term activity could reveal important information about the physical origin of these events. For example, they could be the result of a star-planet interaction (SPI): magnetospheric interactions between planetary and stellar magnetic fields can generate reconnection events and perturbations of the magnetic field topology. As a result, more flares could manifest in a system with hot Jupiter than in single star of the same age \citep{Cohen2011}. In other studies, \cite{Shkolnik2003,Shkolnik2005,Shkolnik2008}, and \cite{Walker2008} reported evidence of chromospheric activity phased with the planetary orbital motion probed by Ca II H \& K lines. Also, in the X-ray \cite{Pillitteri2015} claim to find evidence of flaring activity in phase with planetary period in the system HD 189733, even if this finding is challenged by \cite{Route2019}.
 Furthermore, flaring activity could produce only at specific active longitudes on the stellar surface, in that case, a periodicity based on stellar rotation is expected.

 In this scenario, the advent of a high-precision time-series photometric mission allows studying with great detail the stellar activity. In particular, the Transiting Exoplanet Survey Satellite \citep[TESS;][]{Ricker2014} produces light curves with relatively short cadence (2 minutes) and duration of $\approx 25$ days, sufficient to both observe the single events in detail and to estimate their rate. 
 There are several techniques in literature used to determine the short-term activity of the stars. Most approaches are based on smoothing the light curve and detecting the points above certain criteria \citep{Davenport2014,Stelzer2016,Gilbert2021,Martioli2021}.
 These techniques usually are easy to implement but require often user intervention. Another kind of approach uses deep learning methods \citep[e.g.][]{Vida2021}. These methods, if well trained, are very powerful and do not require any user input. However, the training of neural networks usually is time-consuming. 

 In this work, we present a new automatic procedure to identify flares from any light curve and its application to the first case DS Tucanae A (DS Tuc A). Our approach includes the use of Gaussian processes to remove the stellar long-term contribution to the activity and then a procedure to identify and fit the short-term energetic phenomena, determining their amplitude, decay time, and energy.

 The paper is structured as follows: Sect. \ref{sec:2} describes the procedure used to remove the long-medium activity components in stellar light curves in order to identify and characterize the short-term  phenomena. In Sect. \ref{sec:3}, we present the results obtained from the analysis for DS Tuc A. In Sect. \ref{sec:4}, we discuss the validation of our procedure and a comparison between DS Tuc A and AU Mic. Finally, in Sect. \ref{sec:5}, we drawn our conclusions.  

\section{Data Analysis}\label{sec:2}
In this section, we present our analysis aiming to remove the long-medium components of stellar activity to automatically identify the energetic events at a short timescale.
We analyse DS Tuc A,  a relatively young (age around 40 Myr, \citealt{ Kraus2014,Bell2015,Crundall2019,Benatti2021}) G6V star belonging to the physical binary system DS Tuc, with a light curve  characterized by long-term photometric variations due to a spot cycle \citep{Benatti2021}. DS Tuc A hosts an inflated Hot Jupiter \citep{Benatti2021}. The target is of particular interest since low-density atmospheres are very sensitive to high energy radiation-induced evaporation and photochemistry. We, also, analyse AU Microscopii (AU Mic) an M1 star with an age of 22 Myr \citep{Mamajek2014} that shows an edge-on debris disk and two Neptune-size planets. AU Mic is know to be a magnetically active star with strong flaring activity \citep{Robinson2001}. Its activity is mainly dominated by starspots that produce BY Draconis-type light curve with a quasi-periodic rotational modulation and a period of $4.863\pm 0.010$d \citep{Plavchan2020}. We use this system to validate our technique with the previous analyses already presented in literature \citep[e.g.][]{Martioli2021,Gilbert2021}.

We use all the three 2-minutes cadence TESS light curves available for the system DS Tuc A and the two analogous TESS light curves available for AU Mic\footnote{To be consistent in the comparison with DS Tuc A we used the 2-minutes cadence available data for Sector 1 and 27 for AU Mic.}. We use the data corrected for time-correlated instrumental signatures \citep{Jenkins2016}\footnote{PDCSAP\_SAP flux column in the FITS file.}. Moreover, TESS light-curve FITS files contain a flag that indicates the goodness of the data. For this analysis, we used only data flagged as not polluted by any anomalies (i.e. FLAG$=0$). 
Both DS Tuc A and AU Mic systems are known to host planets. Planetary transits are quite difficult to model correctly and a bad model could result in false flares in the analysis. In order to avoid spurious effects due to transit, before proceeding with the analysis we remove from all the light curves used in this work the bins of the light curves corresponding to the transits of all the known planets for both systems \citep{Benatti2021,Martioli2021}. We consider for the transit time duration the value defined in literature. We, removed the $10\%$ more of the transit duration ($5\%$ before and $5\%$ after the transit), as safety factor to compensate any uncertainties in transit or planetary period.

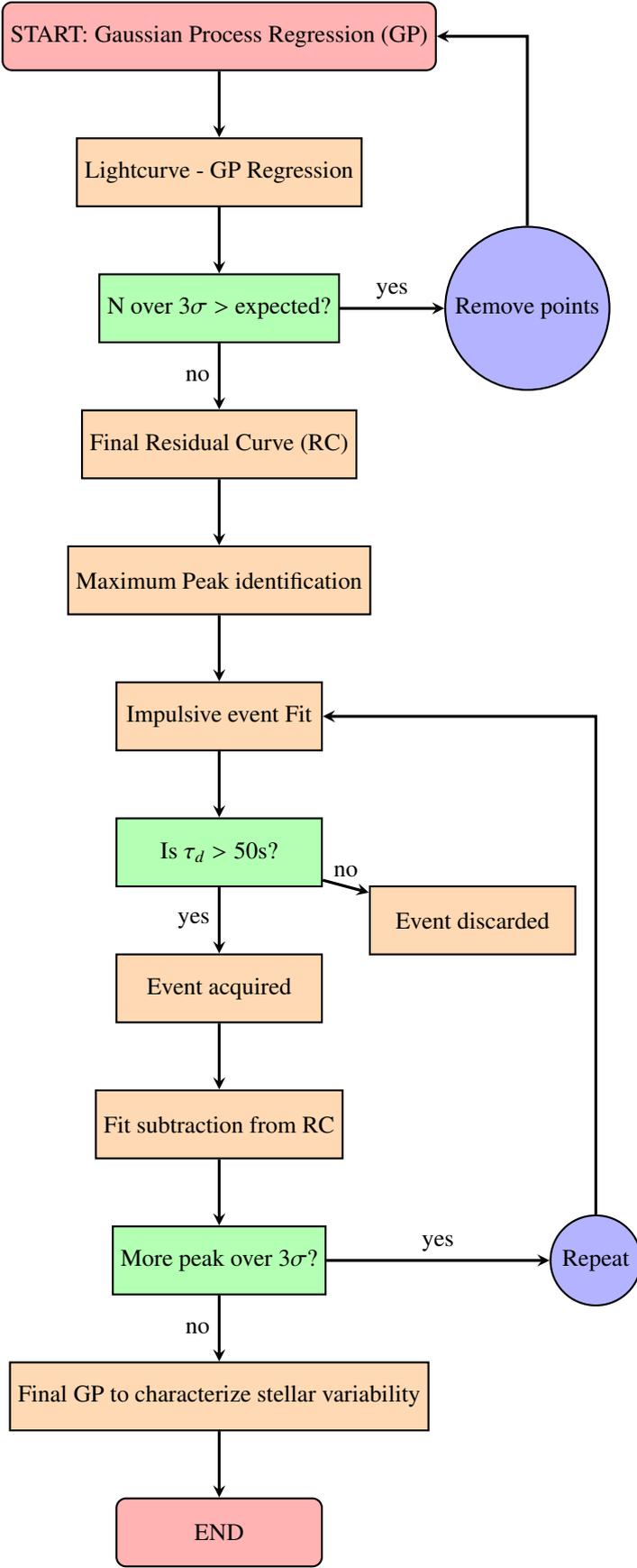
\begin{figure}
\begin{tikzpicture}[node distance = 2.0cm]
\centering
\node (start) [startstop, text centered] { START: Gaussian Process  Regression (GP)};
\node (diff) [process, below of =start] {Lightcurve - GP Regression};
\node (if) [decision, below of = diff]{N over $3\sigma >$ expected?};
\node (rep) [repeat, right of =if, xshift = 2.5cm] {Remove points};
\node (rc) [process, below of =if] {Final Residual Curve (RC)};
\node (peak) [process, below of = rc] {Maximum Peak identification};
\node (fit) [process, below of = peak] {Impulsive event Fit};
\node (qst) [decision, below of = fit] {Is $\tau_d> 50$s?};
\node (qst_yes) [process, below of = qst] {Event acquired};
\node (qst_no)[process, right of =qst,xshift = 1.7cm, yshift=-1.0cm] {Event discarded};
\node (sub) [process, below of = qst_yes] {Fit subtraction from RC};
\node (if peak) [decision, below of = sub]{More peak over 3$\sigma$?};
\node (rep peak) [repeat, right of =if peak, xshift = 3.5cm] {Repeat};
\node (lastgp) [process, below of = if peak]{Final GP to characterize stellar variability};
\node (end) [startstop, text centered, below of = lastgp] {END};

\draw [arrow](start) -- (diff);
\draw [arrow](diff) -- (if);
\draw [arrow](if) -- node[anchor = south]{yes} (rep);
\draw [arrow](rep) |- (start);
\draw [arrow](if) -- node[anchor = east]{no} (rc);
\draw [arrow](rc) -- (peak);
\draw [arrow](peak)--(fit);
\draw [arrow](fit)--(qst);
\draw [arrow](qst) -- node[anchor = east]{yes}(qst_yes);
\draw [arrow](qst) -- node[anchor = south]{no} (qst_no);
\draw [arrow](qst_yes)--(sub);
\draw [arrow](sub) -- (if peak);
\draw [arrow](if peak) -- node[anchor = south]{yes} (rep peak);
\draw [arrow](if peak) -- node[anchor = east]{no}(lastgp);
\draw [arrow](rep peak) |- (fit);
\draw [arrow](lastgp) --(end);
\end{tikzpicture}
\caption{Flow chart for the analysis procedure developed.}
\label{fig:flow_chart}
\end{figure}

To identify the flares we perform a multi-step procedure summarized in Fig. \ref{fig:flow_chart} and described in detail in the following. 

\subsection{Gaussian Process Regression}
Data have been fitted with a Gaussian Processes model (GP) through the python package {\tt celerite} \citep{celerite}, combined to {\tt emcee} \citep{Foreman2013} as sampler. The covariance is described in terms of a stochastically-driven, damped harmonic oscillator, called Simplified Harmonic Oscillator (SHO), of the following form

\begin{equation}
S(\omega) = \sqrt{\frac{2}{\pi}} \frac{\sigma^2/\omega_0^2}{2(\omega^2-\omega_0^2)^2/\tau+\omega_0\omega^2} 
\end{equation}

where $S(\omega)$ is the power spectrum density associated with the frequency $\omega$, $\sigma$ is the standard deviation of the process, $\tau$ is the timescale of the process, $\omega_0$ is the non-damped frequency of the oscillator. An extra error term $\sigma_{Jit}$ is added to the data and included in the fitting procedure to account for unknown sources of error, and it is added in quadrature to data errors. 

Although it has been shown that this kernel function could give non-reliable results for our specific sample, i.e. spot-dominated activity \citep{Perger2021}, with this fitting procedure, we aim to disentangle the long timescale activity from the short timescale signals, in which the flaring activity lives together with the noise, and we do not need to retrieve any physical information at this stage. Therefore, we do not need to run the chain to convergence and we set the chain to run for 10K steps (using 32 walkers). To remove the low probability points due to the random walkers initialization, we set an initial burn-in of 50K steps.

Moreover, we have excluded the short $\tau$ from the parameter space that allows the fit to describe short timescale variability. Consequently, we set the $\tau$ prior to cover only values greater than $10^{3}$ days that guarantee that only the long term variability is described. Details of the imposed priors are presented in Tab. \ref{tab:priors}. 

\begin{table}[t]
\centering
\caption{Model priors. Labels $\mathcal{L}$ $\mathcal{U}$ represent log uniform distribution.}
\label{tab:priors}
\begin{tabular}{l l c}
\hline
\hline
\noalign{\smallskip}
Parameter & Prior & Description \\
\noalign{\smallskip}	
\hline	
\noalign{\smallskip}
\multicolumn{3}{c}{\textit{SHO Model parameters}} \\
\noalign{\smallskip}		
$\sigma$	 & $\mathcal{L} \mathcal{U}$($10^{-5}$, $10^5$)  & \\
$\tau$	&  $\mathcal{L} \mathcal{U}$($10^3$, $10^4$) & (d)\\
$\omega_0$ & $\mathcal{L} \mathcal{U}$(2$\pi$/50, 2$\pi$)& (1/d)\\
$\sigma_{Jit}$ & $\mathcal{L} \mathcal{U}$($10^{-2}$, $10^2$)& (e$^{-}$/s)\\

\multicolumn{3}{c}{\textit{QP Model parameters}} \\
$h$	 & $\mathcal{L} \mathcal{U}$($10^{-2}$, $10^2$)  & \\
$\omega$	&  $\mathcal{L} \mathcal{U}$ ($10^{-2}$, $10$) & (1/d)\\
$\tau$ & $\mathcal{L} \mathcal{U}$($10^{-1}$, $10^3$)& (d)\\
$P$ & $\mathcal{L} \mathcal{U}$($1$, $10^1$)& (d)\\
\noalign{\smallskip}							    
\hline	
\noalign{\smallskip}
\end{tabular}
\end{table}			

\begin{figure}
    \centering
    \includegraphics[scale = 0.5]{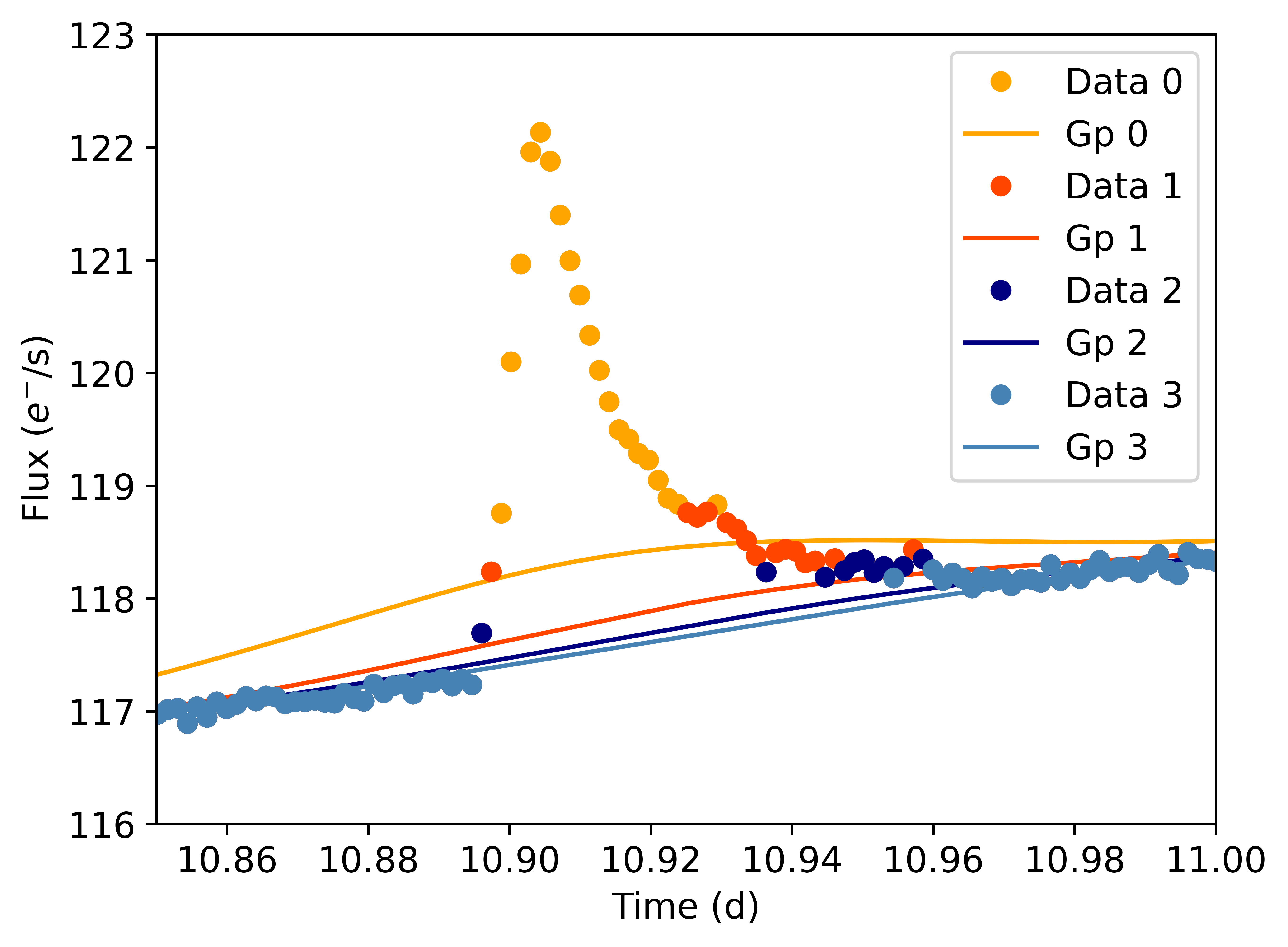}
    \caption{Time vs flux zoom showing the iterative GP procedure close to a short term event. Each curve shows different steps of the fitting procedure.}
    \label{fig:gp_method}
\end{figure}

However, the presence of the Jitter in the fitting procedure allows the GP to describe a small fraction of the short timescale signals. To avoid this, we iterate the GP fit removing at each step the data over 3 sigmas from the residual curve, obtained by the subtraction from data analysed at step $n$ of the maximum a-posteriori probability (hereafter MAP) GP prediction obtained in the $n_{th}$ iteration (See Fig. \ref{fig:flow_chart}).
This expedient removes the data around a peak and the GP, consequently, smooths out the data ignoring the short timescale signals. The process is repeated until the number of data points over 3 sigmas is compatible with what we expected from the noise.
An example of the procedure is presented in Fig.\ref{fig:gp_method}, in which data points are presented together with the GP prediction, at different iterations.

\begin{figure*}[!h]
    \centering
    \subfigure[]{\includegraphics[scale = 0.35]{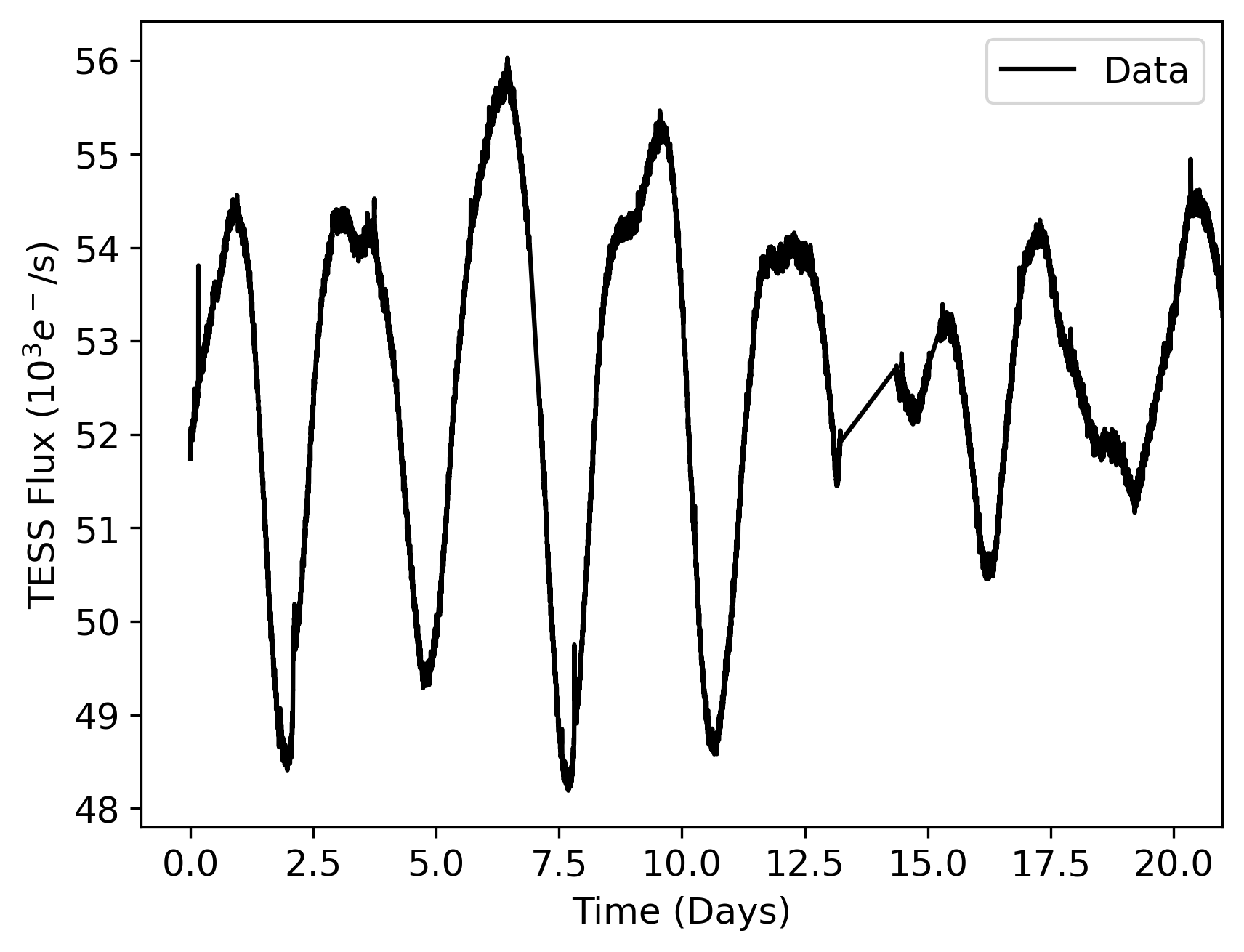}} \subfigure[]{\includegraphics[scale = 0.35]{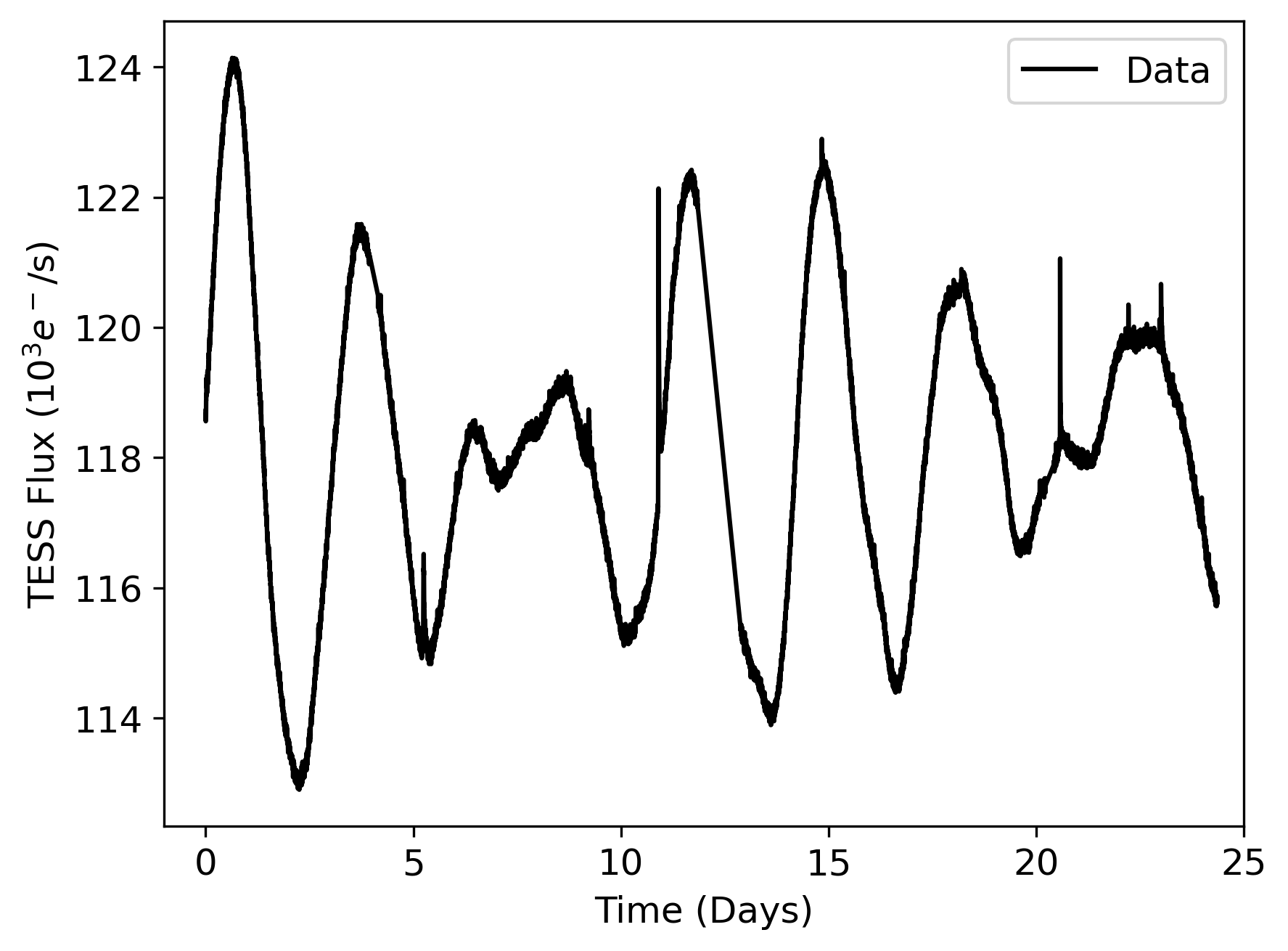}}
    \subfigure[]{\includegraphics[scale = 0.35]{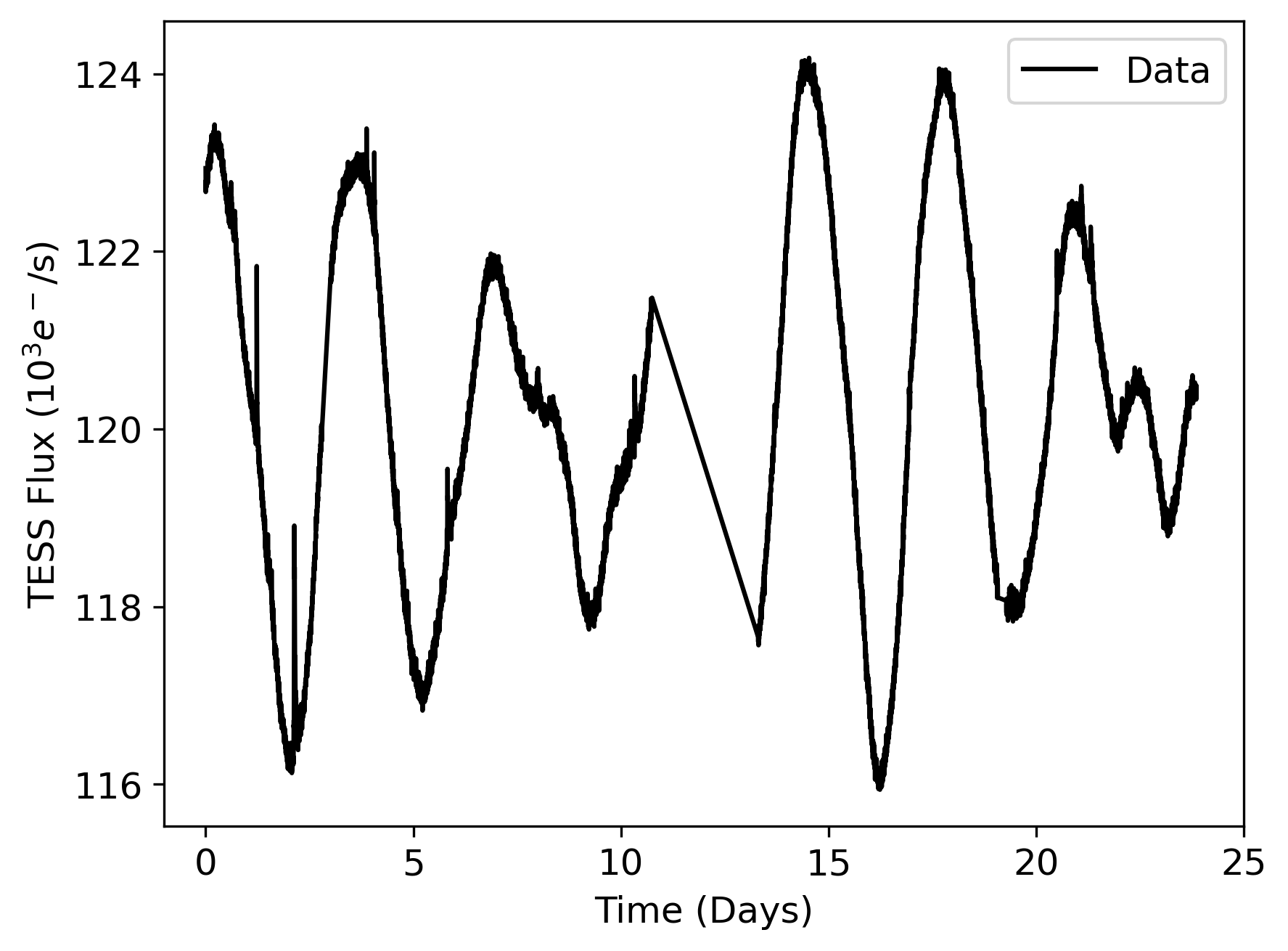}} 
    \centering
    \subfigure[]{\includegraphics[scale = 0.35]{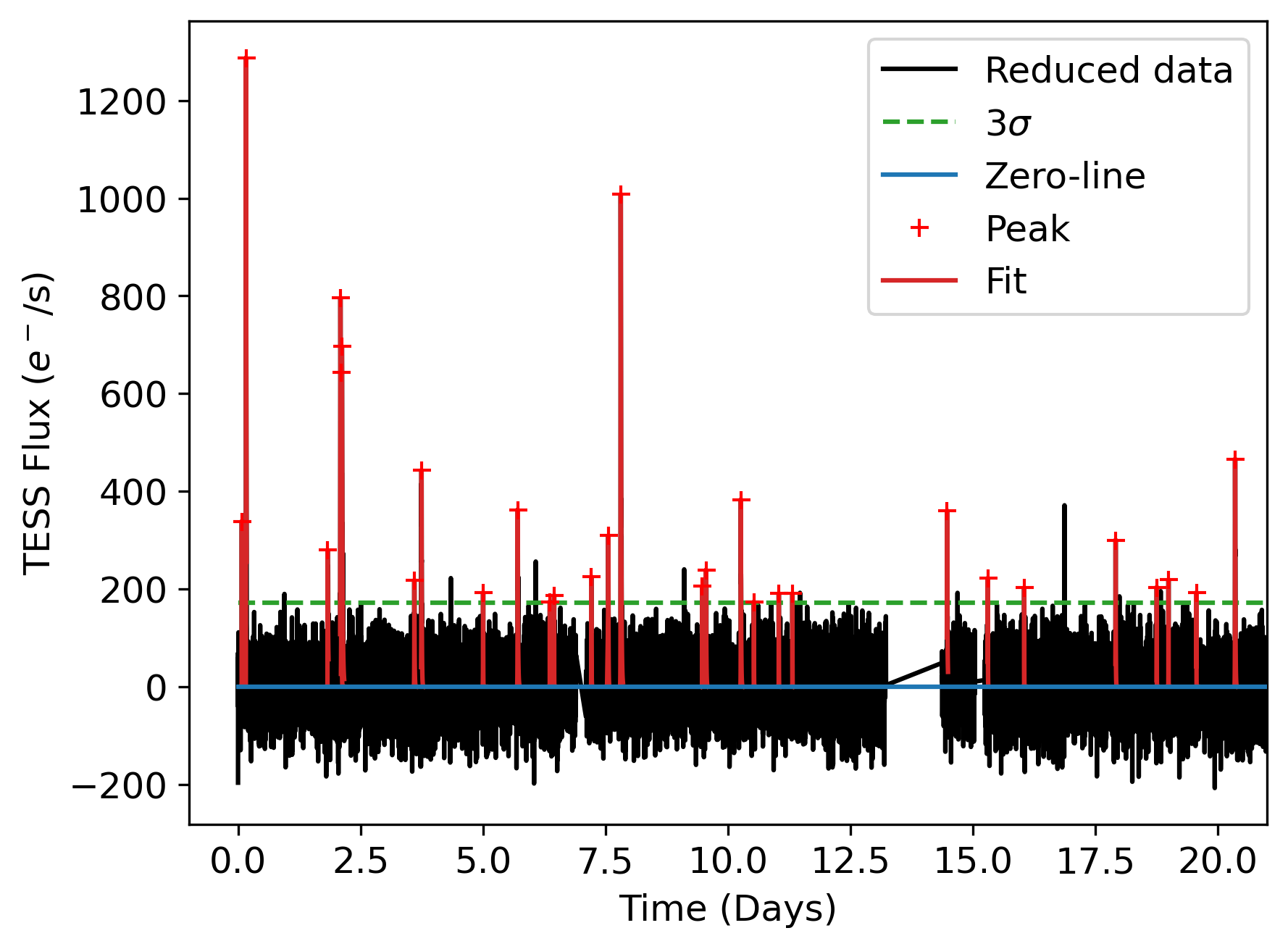}} \subfigure[]{\includegraphics[scale = 0.35]{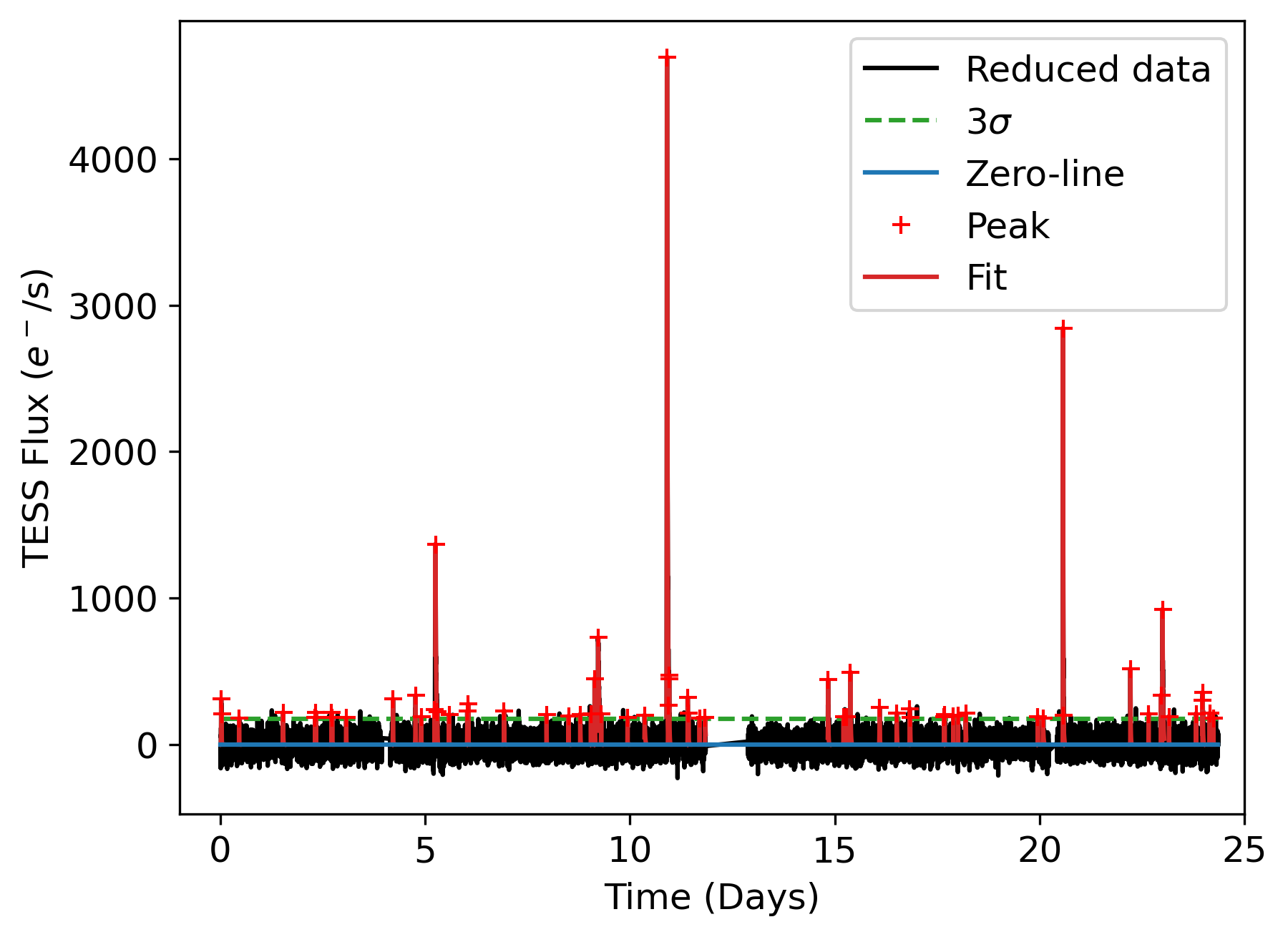}}
    \subfigure[]{\includegraphics[scale = 0.35]{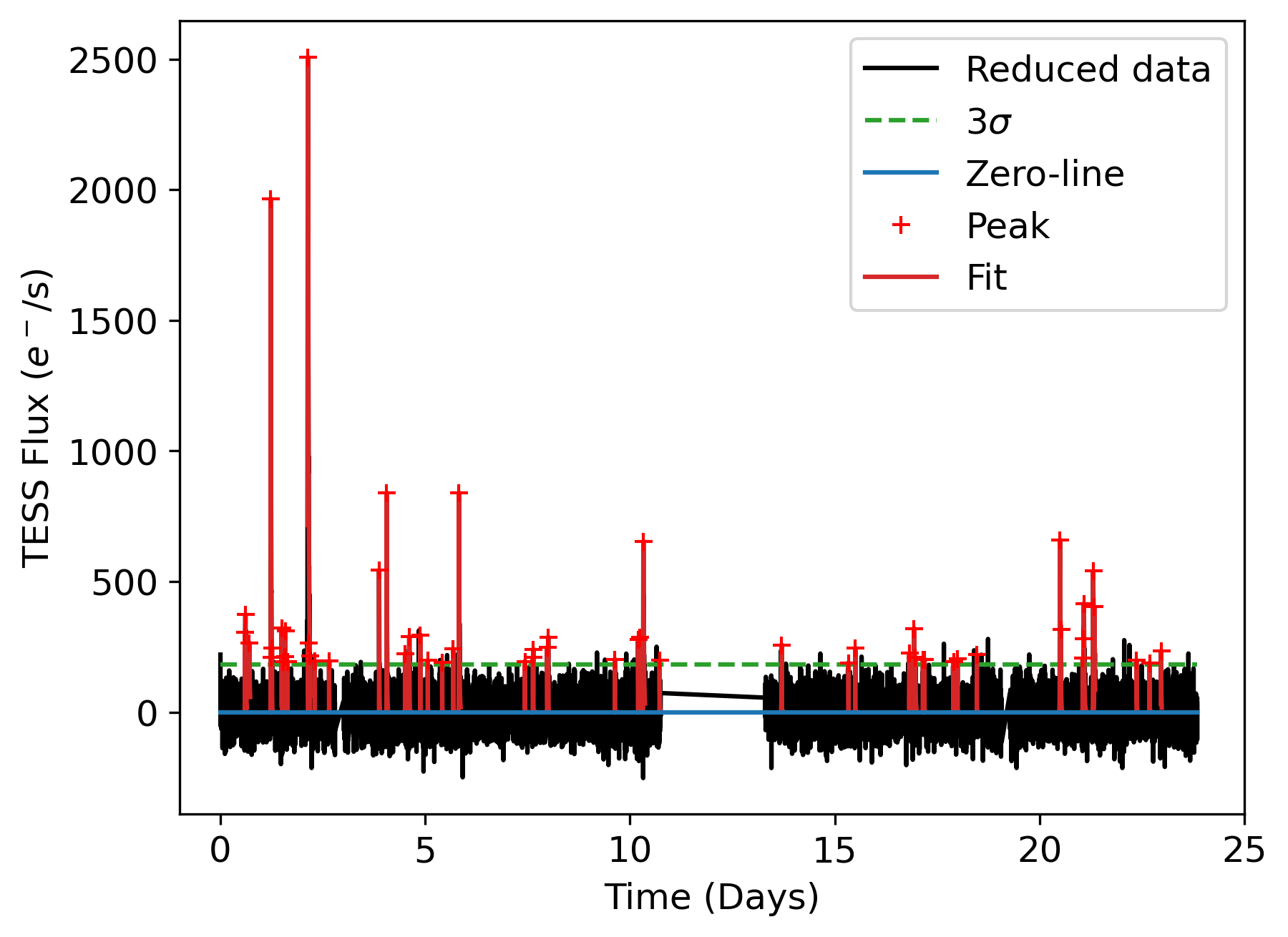}}

    \caption{The first step of the analysis. On the top panels (a,b,c) the three initial light curves (Sector 0, 27 and 28 respectively). On the bottom panel (d,e,f) are showed the final RCs after the GP analysis. The red crosses show the peaks of each event identified and the red lines the fits for each event. The dotted green line shows the $3\sigma$ value and the blue line is the zero-line. }
    \label{fig:lc_gp}
\end{figure*}

 The subtraction from the initial data of the last MAP GP prediction gives us a final residual curve (RC) centered at zero, in which all the small scale events (the points removed in the iterative process) are highlighted. This is presented in Fig. \ref{fig:lc_gp}, where the DS Tuc A light curves and their resulting final residual curves are shown together with the identified events (red lines).

\subsection{Events identification and characterization }
\begin{figure}[!h]
    \centering
    \includegraphics[scale =0.5]{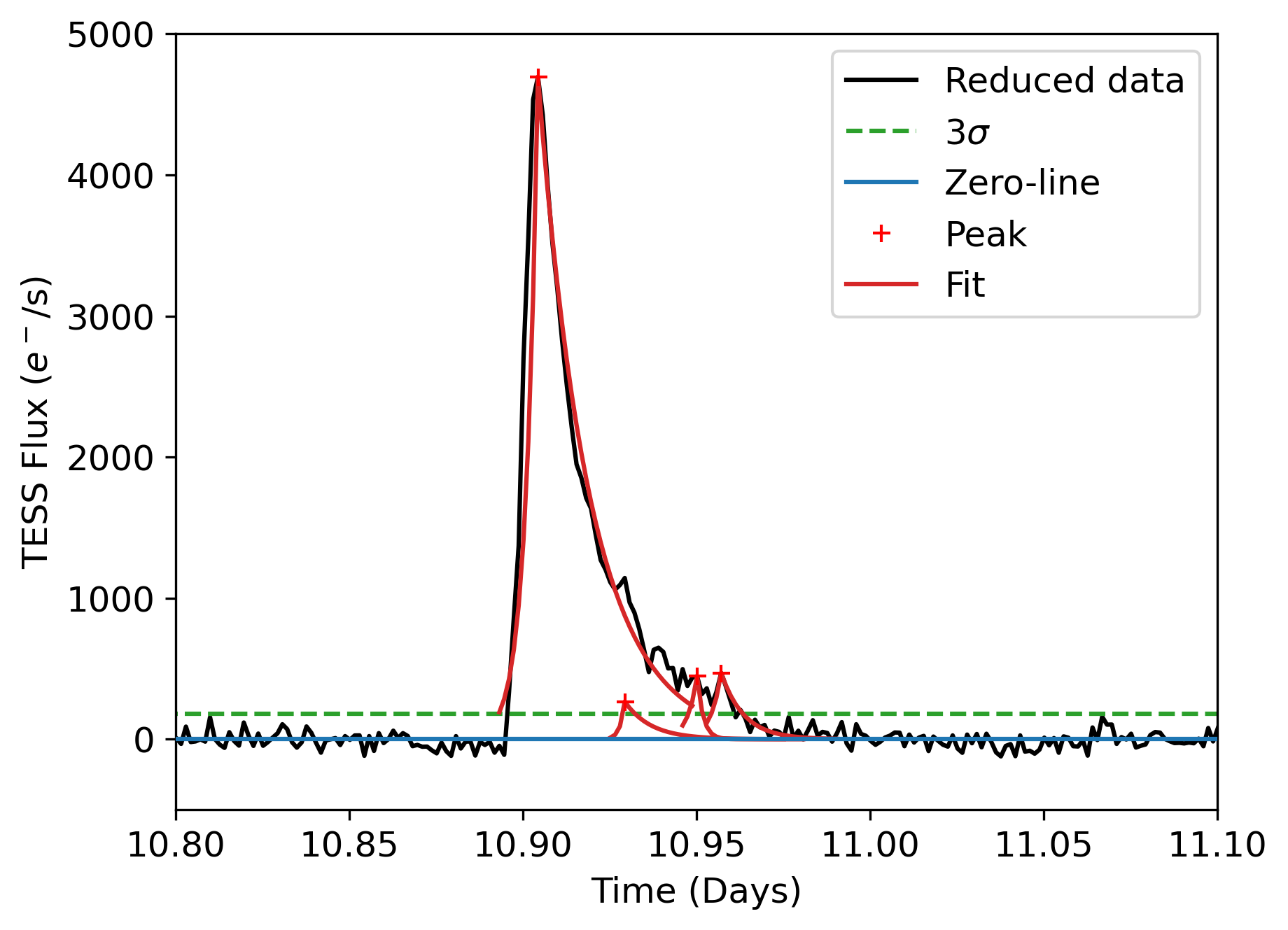}
    \caption{Close-up view on one multiple-event identified. Legend as in Fig. \ref{fig:lc_gp}. }
    \label{fig:flare}
\end{figure}

The next step in our analysis identifies and fits the short-term activity events from the RC (See Fig. \ref{fig:flow_chart}). 
Then, we identify the maximum between all the local curve maxima above $3\sigma$. The code then fit the events using a function composed of an exponential increase and an exponential decay with maximum the event peak $A$ located at $t_0$ (the time coordinate of the peak):
\begin{equation}
    F(t)= H(t_0-t)\cdot A \exp \left({\frac{t-t_0}{t_{r}}}\right) +     H(t-t_0)\cdot A \exp\left({\frac{t_0-t}{t_{d}}}\right)
    \label{eq:flare}
\end{equation}
where we use $H(t_0-t)$ the Heaviside function (the value of which is zero for negative arguments and one for positive arguments) to switch from the rising part to the decaying part of the function.
Then, we subtract the fit to the RC curve in order to highlight smaller energetic events occurring at the same time that will be identified through an iterative procedure. Fig. \ref{fig:flare} shows an example of fitting on the RC where multiple events are present. We stress that our goal is to describe the short term variability, not to derive the physics of each event, therefore, even a unique complex event, will be described by multiple elementary events as in Eq. \ref{eq:flare}.
The code fits all the events until there are no more local maxima over the 3$\sigma$ threshold defined above.
To avoid spurious points to be considered as energetic events, the code keeps an event only if the associated decaying time is greater than 50 s, otherwise, the event is discarded from further analysis.   

\subsection {\bf Final products}
After the analysis, we obtain for each light curve the number of identified short term events, their amplitudes and positions in time, their rising and decaying times and their energies (See Sect. \ref{sec:3}).
After that, we remove the energetic impulsive events from the original light curve and we run again a GP on the residual data (without the short term events) in order to better constrain the stellar activity parameters, i.e. the rotational period and the timescale of the active region lifetime.

For this task, we select a different kernel, i.e. the quasi-periodic kernel within {\tt george} python package \citep{hodlr}, whose expression is the following

\begin{equation} \label{eq1}
\begin{split}
k(i,j) & = h^2 \exp{\left(-\frac{(t_i-t_j)^2}{\tau^2} -\frac{\sin^2(\pi(t_i-t_j)/P_{rot})}{2\omega^2}\right)} \quad + \\
 & + \delta_{ij} (\sigma_{TESS}^2+\sigma_{Jit}^2)
\end{split}
\end{equation}

where $k(i,j)$ is the $ij$ element of the covariance matrix, $t_i$ and $t_j$ are two times of the light curve data set, $h$ is the amplitude of the covariance, $\tau$ is the timescale of the exponential component (active region lifetime), $\omega$ is the weight of the periodic component (shape of the periodic component), $P_{rot}$ is the (rotational) period, $\delta_{ij}$ is the Dirac delta function, $\sigma_{TESS}$ are the data errors and $\sigma_{Jit}$ is a white noise term added to account for unknown source of uncertainties. This kernel has been proved to gives reliable results even in the case of stars with spots dominated activity \citep{Perger2021}.
For this task, we run the chain for 10K steps after a 50k steps burn used to initialize the 32 random walkers.
 We decided to adopt in this phase a different kernel since now our goal is to derive parameters linked to the physical properties of the medium-term variability of the star, while in the previous step we aimed simply at a description of the shape of the stellar background. In that phase, we chose a less computing expensive kernel, probably less representative of the physical origin of the variability, because of the shorter computation time and the need to proceed with an iterative procedure.

A rebinning procedure has been performed on the light curves in order to reduce the computational cost. The bin step has been set to 1 over 24 hours in which we have averaged both times and fluxes. We have calculated each bin error as 

\begin{equation}
    err = \sum_{i=1}^N (yerr_i)/N
\end{equation}

where $yerr_i$ is a flux error and $N$ is the data count which falls in the bin. We neglect the empty bins.
The results of this analysis for the 3 sectors of DS Tuc A and the 2 sectors of AU Mic are summarised in Tab. \ref{tab:last_gp}. 

\begin{table*}[]
    \caption{Results of the GP for the long term activity. $\sigma_{Jit}$ is omitted because is always lower than $10^{-4}$.}
    \label{tab:last_gp}
    \centering
    \begin{tabular}{c c c c c}
    \hline
    \hline
         Star   & $h$ &  $\omega$ & $\tau$  &    $P$  \\
    \hline
         DS Tuc Sector 1  & ${0.031}_{-0.004}^{+0.006}$   &  ${0.51}_{-0.05}^{+0.06}$ &  ${2.8}_{-0.3}^{+0.3}$ & ${3.02}_{-0.04}^{+0.04}$  \\
         DS Tuc Sector 27  & ${0.0172}_{-0.0019}^{+0.0024}$   &  ${0.48}_{-0.03}^{+0.04}$ &  ${3.5}_{-0.3}^{+0.3}$ & ${2.97}_{-0.04}^{+0.03}$  \\
         DS Tuc Sector 28  & ${0.016}_{-0.002}^{+0.003}$   &  ${0.49}_{-0.04}^{+0.05}$ &  ${3.9}_{-0.3}^{+0.3}$ & ${2.93}_{-0.02}^{+0.02}$  \\
         AU Mic Sector 1  & ${0.0110}_{-0.0007}^{+0.0010}$   &  ${0.205}_{-0.009}^{+0.010}$ &  ${8.0}_{-0.3}^{+0.3}$ & ${4.64}_{-0.011}^{+0.012}$  \\
         AU Mic Sector 27  & ${0.0104}_{-0.0003}^{+0.0007}$   &  ${0.281}_{-0.014}^{+0.015}$ &  ${8.3}_{-0.4}^{+0.4}$ & ${4.94}_{-0.03}^{+0.02}$  \\
    \hline
    \end{tabular}
    \label{tab:fit}
\end{table*}


\section{Light curve analysis} \label{sec:3}
TESS data of DS Tuc A covers about 70 days, divided into three different sets, all analysed in this work. 
Table \ref{tab:obs} reports the dates of the relevant sectors. 
\begin{table}[!h]
    \caption{Observations data.}
    \centering
    \begin{tabular}{c c c c}
    \hline
    \hline
    Star   & Sector & Date       & Duration (days) \\
    \hline
    DS Tuc A &   1    & 25-07-2018 &  21.0\\
           &  27    & 05-07-2020 &  24.4\\
           &  28    & 31-07-2020 &  23.9\\
    AU Mic &   1    & 25-07-2018 &  27.1\\
           &  27    & 05-07-2020 &  24.4\\
    \hline
    \end{tabular}
    \label{tab:obs}
\end{table}
The procedure described in the previous section has been applied separately to each sector.  The Sector 1 light curve shows a low signal/noise ratio after 21 days of observations, to avoid spurious points being identified as flares we excluded this part of the curve ($\approx 6$ days) from the analysis.
The light curves and the resulting RCs are reported in Fig. \ref{fig:lc_gp}.

\begin{figure}
    \centering
    \subfigure[][]{\includegraphics[scale = 0.5]{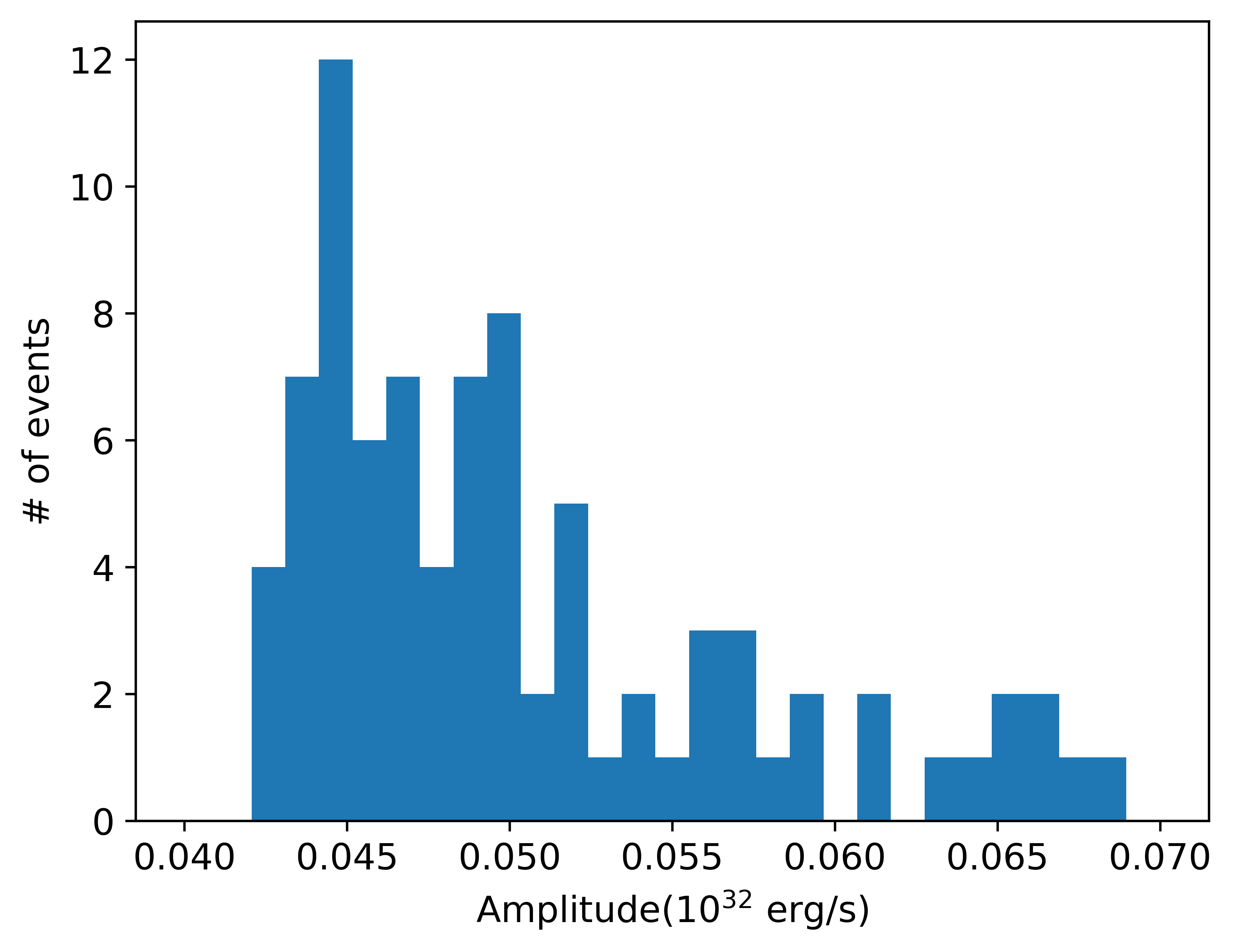}}
    \centering
    \subfigure[][]{\includegraphics[scale = 0.5]{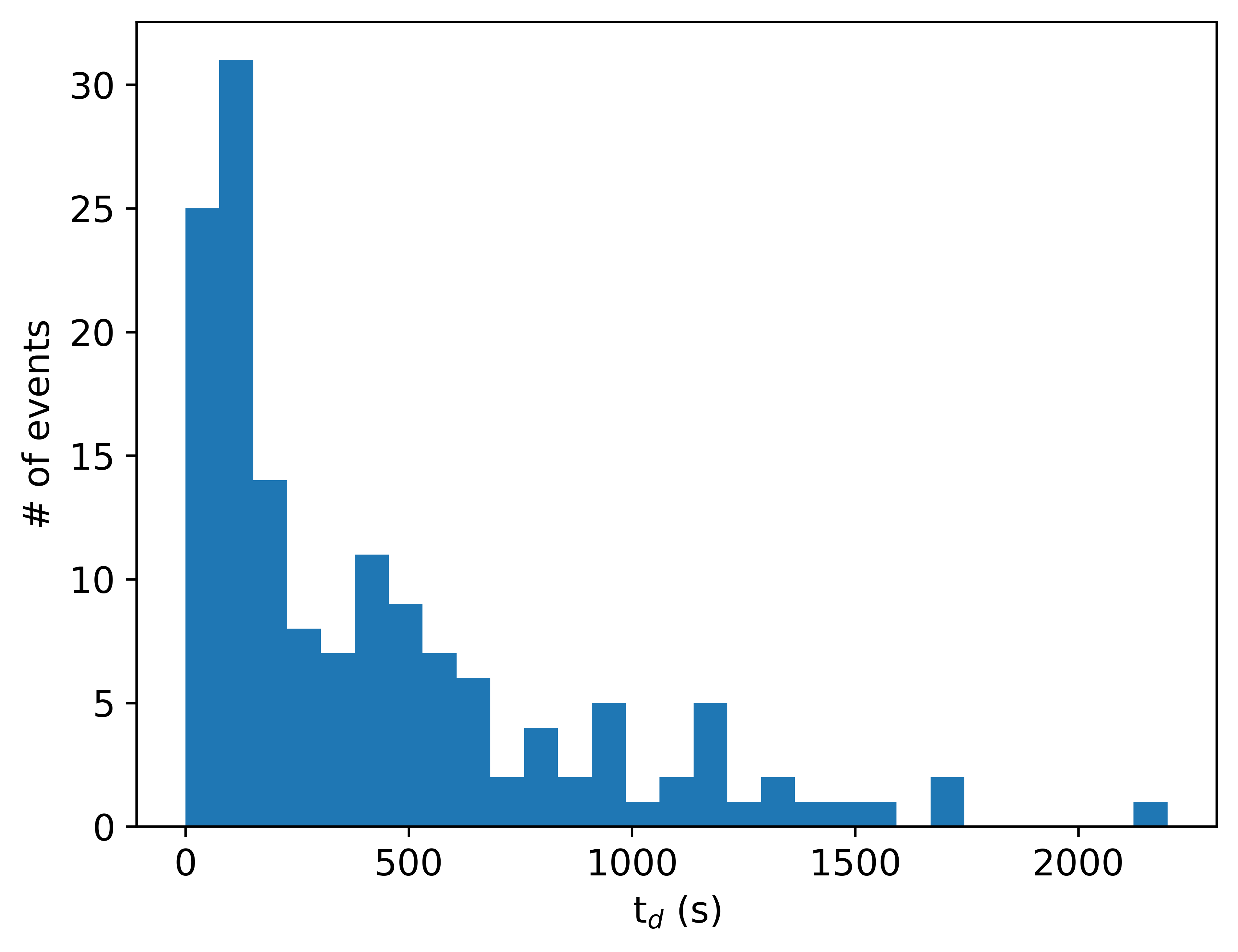}}
    \caption{Top panel (a) histogram for events amplitudes for DS Tuc A. Bottom panel (b) same as top panel for decaying time.}
    \label{fig:hist}
\end{figure}

In particular, we identify 29, 63, 56 events in the three sectors respectively.
Fig. \ref{fig:hist} shows the amplitude and decaying time distributions for the energetic events of DS Tuc.

We estimate the energy associated with each event integrating the two fitted exponentials, this results in:
\begin{equation}
    E = (A\cdot t_r + A\cdot t_d) \frac{L_{bol}}{L_{TESS}}
    \label{eq:energy}
\end{equation}
where $A$ is the event peak, $t_r$ and $t_d$ are the rising and decaying times estimated by the fit, $L_{bol} = 0.725 
L_{\odot}$ is the optical luminosity for DS Tuc A \citep{Newton2019}
and $L_{TESS}$ is the count in the TESS curve\footnote{ We are assuming that the temperature of the flaring region is similar with the photospheric one. This likely is  not correct  producing an underestimation of the true emitted energy but is adequate for our goal and does not affect the slope of the energy distribution.}. In order to associate an error to the energy, for each event we performed a bootstrap procedure with 1000 different fits, where the points on the light curves are taken randomly from a Gaussian with $\sigma$ equals the error associated with the point. The relative error associated to the event energy is given by $\frac{E_{fit} - E_{iter}} {E_{fit}}$, where $E_{fit}$ is the value derived from the first fit and $E_{iter}$ is the average of the energies estimated with the bootstrap procedure. However, with this procedure the associated errors are of the order of 0.1\%.

\begin{figure}[!h]
    \centering
    \includegraphics[scale =0.5]{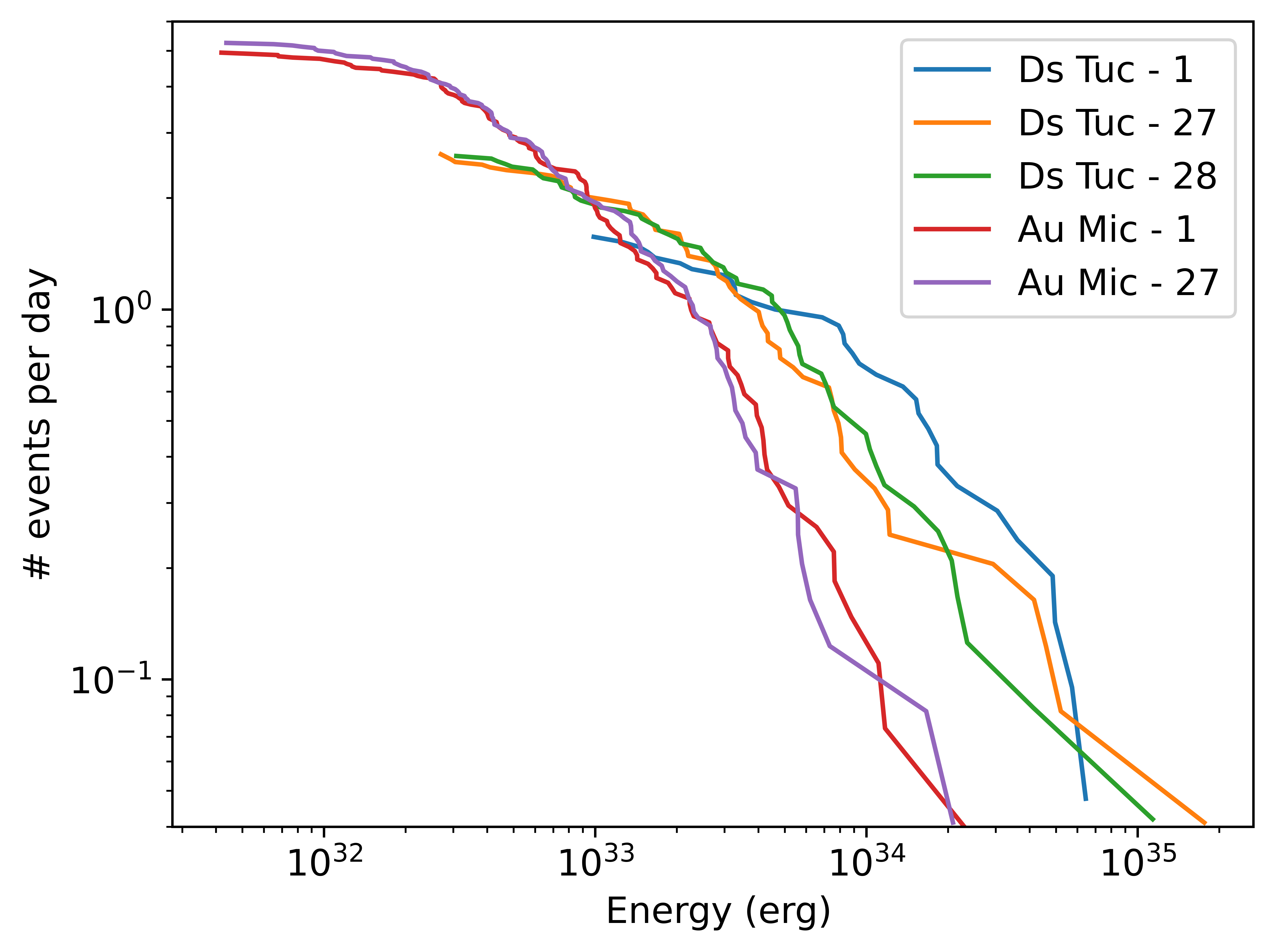}
    \caption{Cumulative events per day vs energy for each set of observations of DS Tuc A and AU Mic.}
    \label{fig:rate}
\end{figure}

Fig. \ref{fig:rate} shows the rate of events energies for the three sectors of DS Tuc A and for the two sectors of AU Mic that will be discussed in the following section.

\begin{figure}
    \centering
    \includegraphics[scale =0.5]{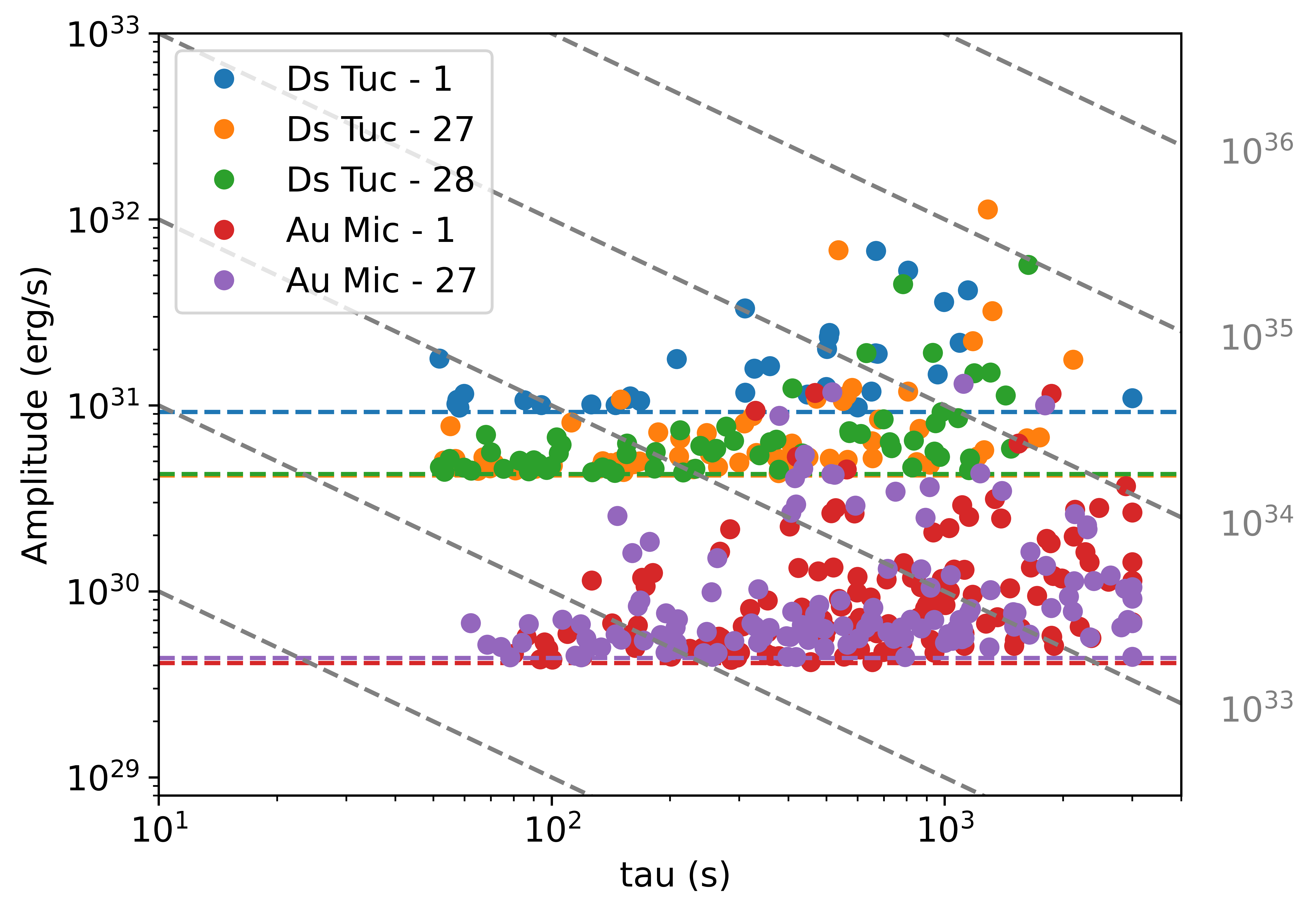}
    \caption{Amplitude vs decaying time for each identified event. The dotted horizontal coloured lines show the 3$\sigma$ value used to discriminate the events from the noise. The dotted grey oblique lines are the iso-energy lines.}
    \label{fig:timevsampl}
\end{figure}

For DS Tuc, we find $\approx 2$ events per day with energy $E> 2\times10^{32}$ erg. However, is clear from the cumulative curve in Fig. \ref{fig:rate} that the sensibility to events starts to increase at energy  $E>2 \times 10^{33}$.
Fig. \ref{fig:timevsampl} reports the amplitude as a function of decay time $t_d$ of the identified events. The figure shows also the thresholds adopted to select the events (that depends on the specific sector) and the curves with constant energy. Our procedure is biased toward high peak-short decay time events and tends to miss long events with a low amplitude of similar energy. In fact, at low energy, the peak of faster events satisfy the criteria to be above the threshold, whereas the peaks of longer events with the same energy are not enough to be over the threshold.

\begin{figure}[!h]
    \centering
    \includegraphics[scale =0.5]{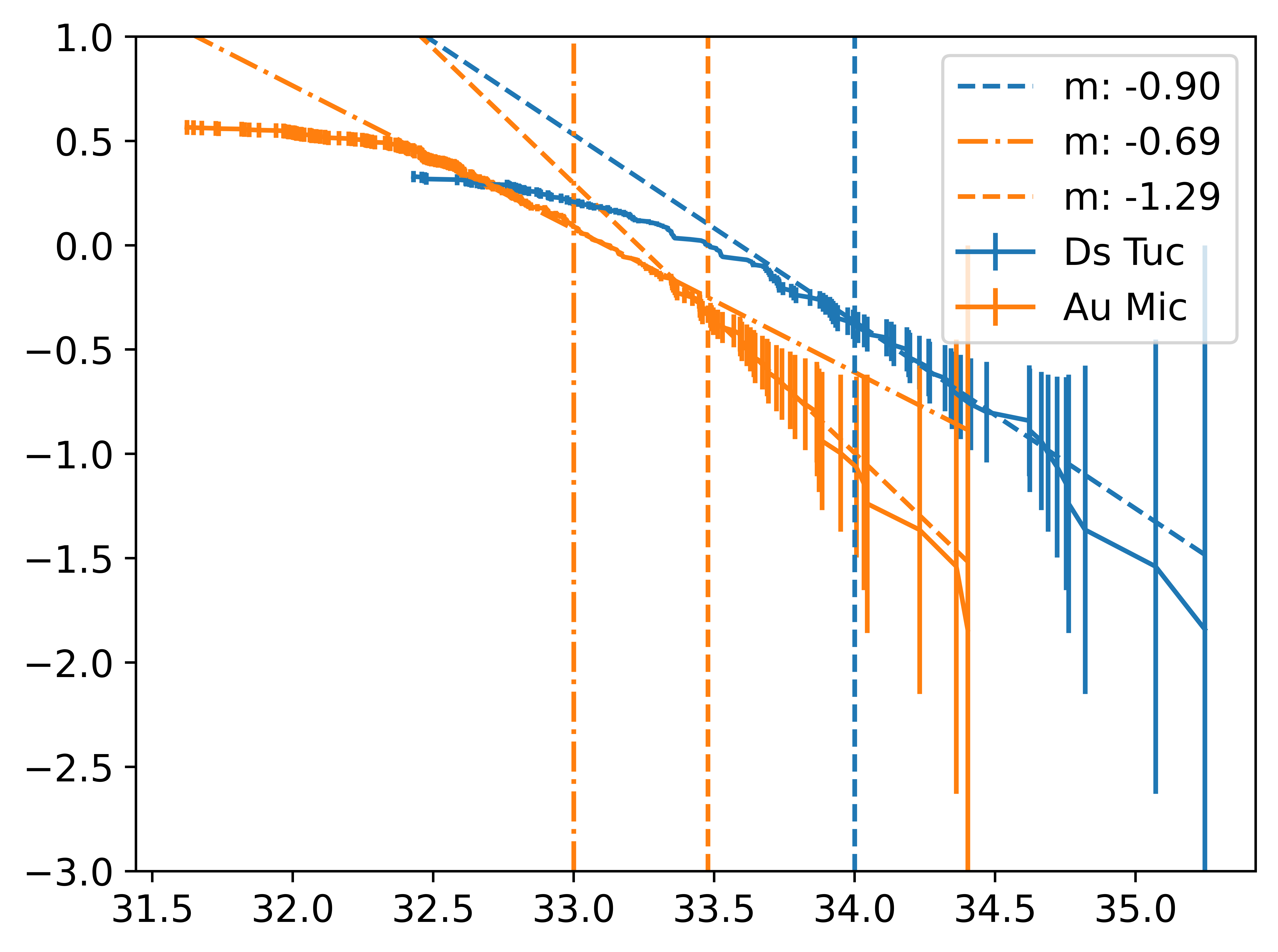}
    \caption{Cumulative log frequency vs log energy for the total set of observations. Error bar are given by the Poisson noise $\sqrt{N}$. The dashed blue line represents the linear fit for Ds Tuc. The yellow lines show the fits for AU Mic at energies higher than $2\times 10^{33}$ erg (dashed line) and for energies higher than $10^{33}$ erg (dashed-dotted line). The vertical lines are located at the break energy used for each fit. }
    \label{fig:cumulative}
\end{figure}

Since the star does not show significant differences in the three sectors we can sum up the individual curves in Fig. \ref{fig:rate} to obtain the cumulative energy frequency distribution of our target (Fig. \ref{fig:cumulative}). As discussed above, the break of the distributions in the region of weak events is a result of the biased procedure. From Fig.\ref{fig:timevsampl} and Fig. \ref{fig:cumulative} we identified the break energy (where the identification of flares is considered complete) as $E_b = 10^{34}$erg for the cumulative curve of DS Tuc A. This choice is supported by Fig. \ref{fig:timevsampl} (and also by Fig.\ref{fig:hist_Ds_Tuc}) that shows that most of flares of DS Tuc have time decay shorter than 2000 s, that for our amplitude threshold corresponds approximately to our break energy (implying that we miss very few - if any - flares above our threshold).
For AU Mic the identification of the break energy is not straightforward, since, as suggested by Fig. \ref{fig:timevsampl}, it shows flares with longer time decay and, likely, we miss several events even at high energies. For this reason, we present the results using two different break energies: $10^{33}$erg and $3\times10^{33}$erg.
As an independent check, we have computed the slope of the curve as a function of the break energy. In the case of DS Tuc A, we find that the slope reach a plateau for break energies between $10^{34}$ erg and $3\times 10^{34}$ erg: for higher break energy the statistics is too low to give robust results, and for lower energy the curve is dominated by incompleteness. On the contrary, for the case of AU Mic the slope does not reach a plateau in any range, confirming that the curve is virtually incomplete throughout the entire high-statistic range. 

In order to estimate the errors in the fitting parameters for both stars taking into account the statistics, we perform a bootstrap procedure.
From the set of $N$ events for each star we create $1000$ new sets. Each of them is made extracting $N$ times a random flare from the original set, it is worth to notice that once that a flare is extracted it is again made available for the following extraction. This procedure takes into account the different frequency of high and low energy flares.
For each set, we calculate a cumulative curve and we perform the fits using the function $f = m\cdot \log(E) +q$ for the part of the curves at energy higher than $E_b$. In this way, for each star we obtain a distribution of $m$ and $q$. The fit results are defined as the mean values for each distribution with the standard deviation as error. 
The cumulative functions with their fits are showed in Fig. \ref{fig:cumulative}. The  parameters of the fits are reported in Tab. \ref{tab:fit}.

\begin{table}[]
    \caption{Results of the linear fits {\bf of the cumulative distributions shown in fig.\ref{fig:cumulative}}.}
    \centering
    \begin{tabular}{c c c c}
    \hline
    \hline
         Star   & $E_b$ (erg) &   $m$  &    $q$  \\
    \hline
         DS Tuc A & $10^{34}$ & $0.90\pm 0.17$ & $30\pm 6$   \\
         AU Mic & $10^{33}$ & $0.69\pm 0.15$ & $23\pm 5$  \\
         AU Mic & $2\times10^{33}$ & $1.29\pm 0.22$ & $43\pm 7$ \\
    \hline
    \end{tabular}
    \label{tab:fit}
\end{table}

\begin{figure*}[!h]
    \centering
    \subfigure[]{\includegraphics[scale =0.5]{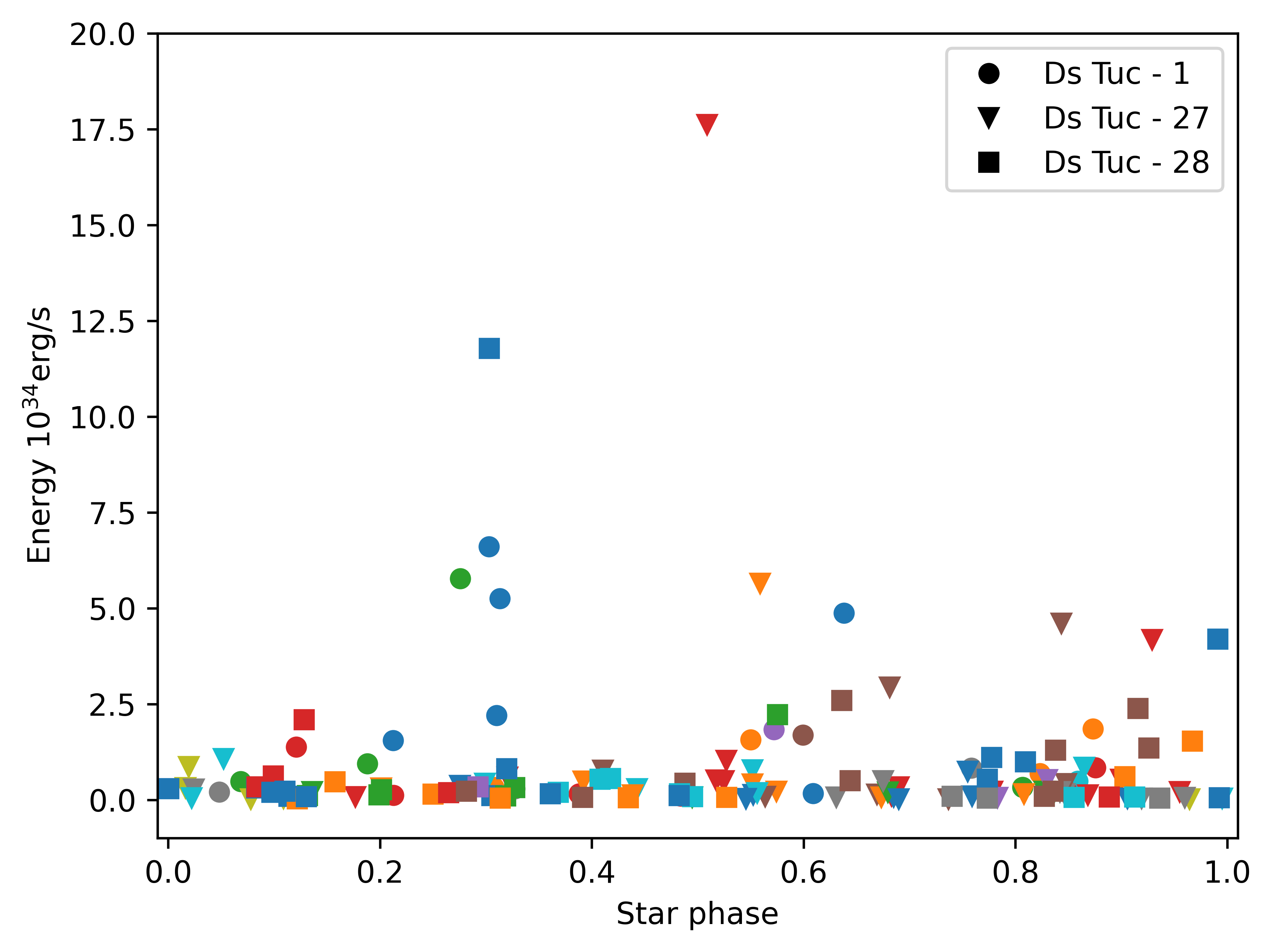}}
    \subfigure[]{\includegraphics[scale =0.5]{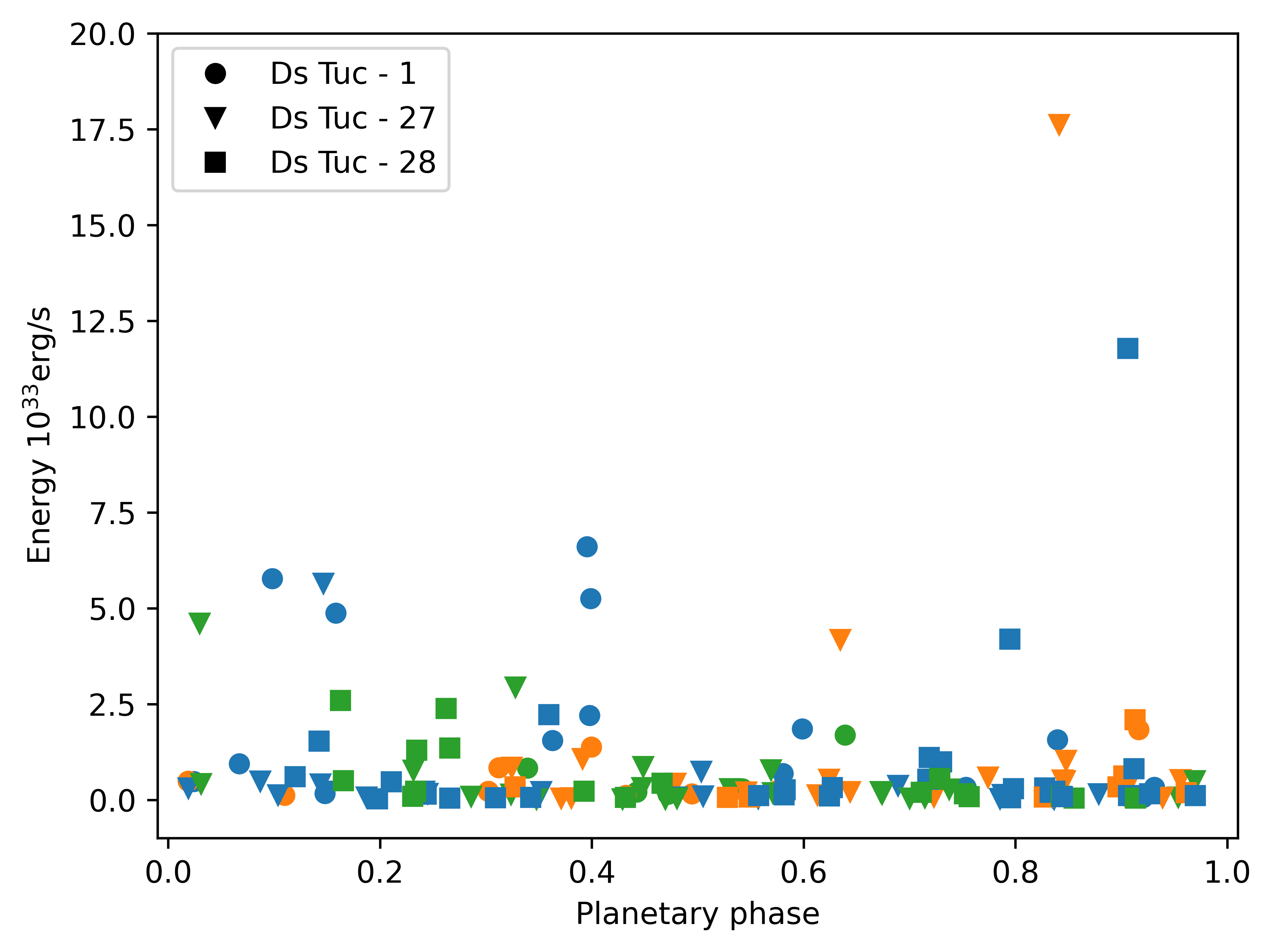}} \\
    \subfigure[]{\includegraphics[scale =0.5]{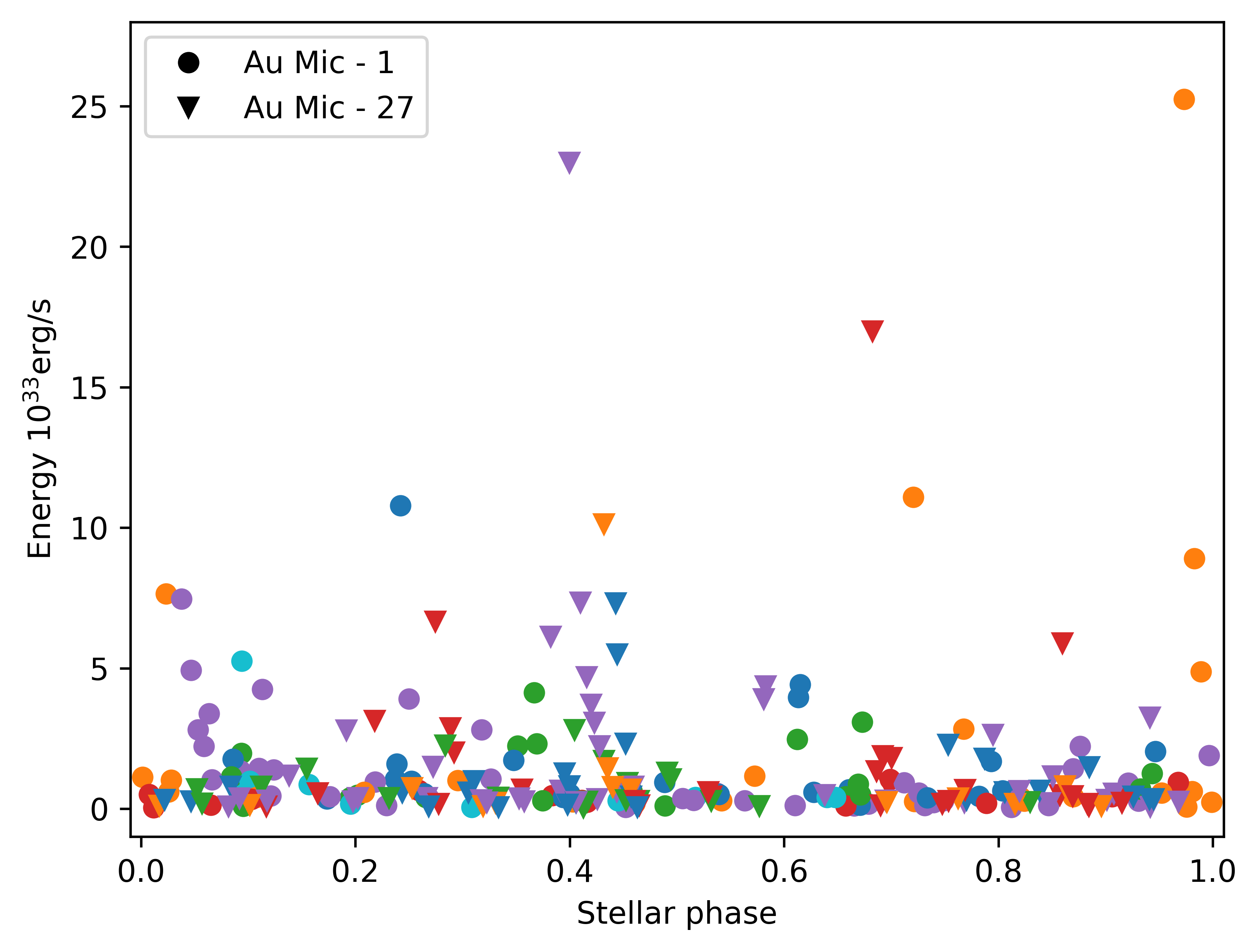}}
    \subfigure[]{\includegraphics[scale =0.5]{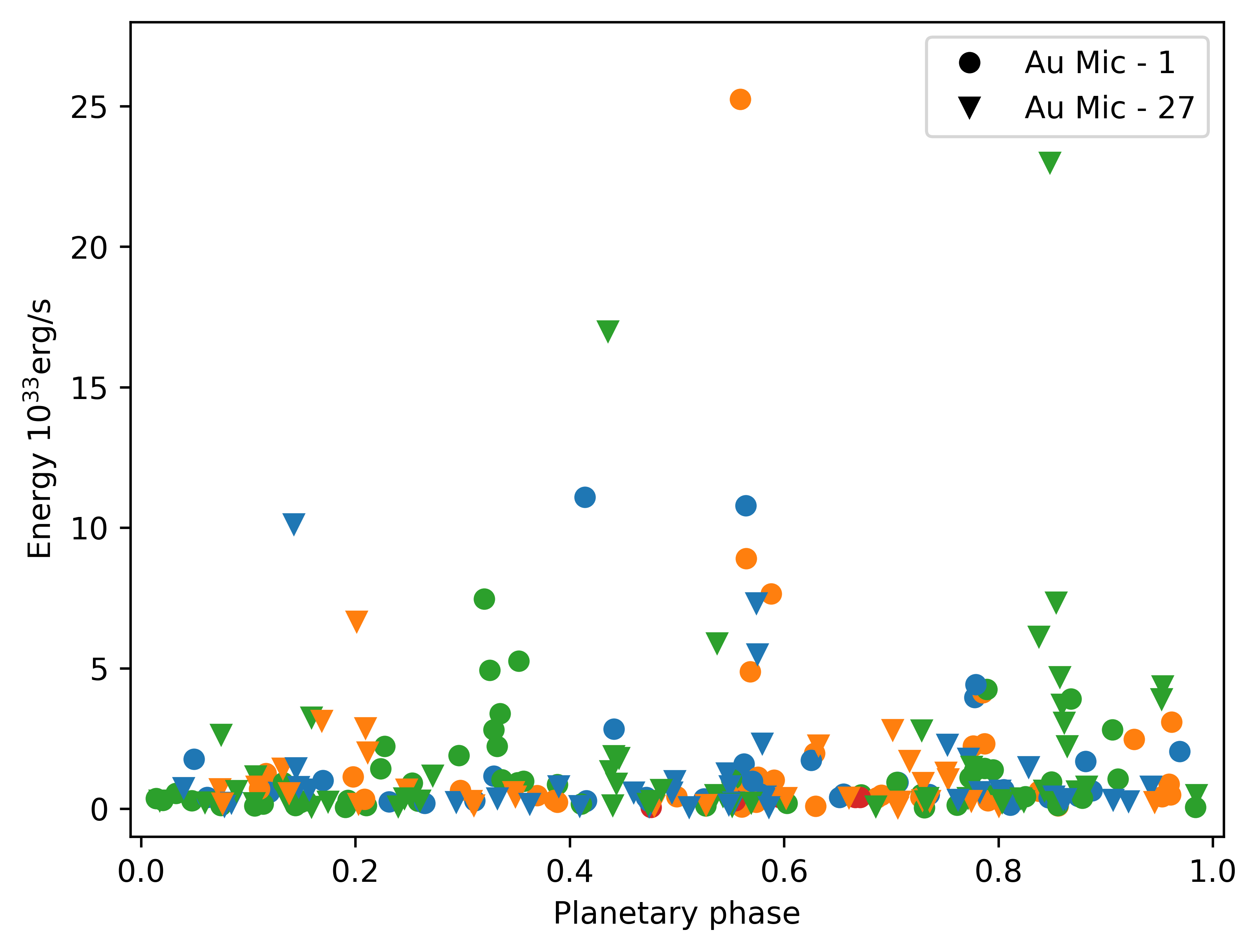}}
    \caption{Top: left panel (a) event energy vs event position phase-folded for the stellar rotation cycles for DS Tuc A. Top right panel (b) same as the left panel but for the planetary period. Bottom panels as top panels but for AU Mic. Different symbols refer to different TESS sectors, while different colors represent different stellar or planetary periods.}
    \label{fig:folding}
\end{figure*}

\begin{figure*}[!h]
    \centering
    \subfigure[]{\includegraphics[scale =0.5]{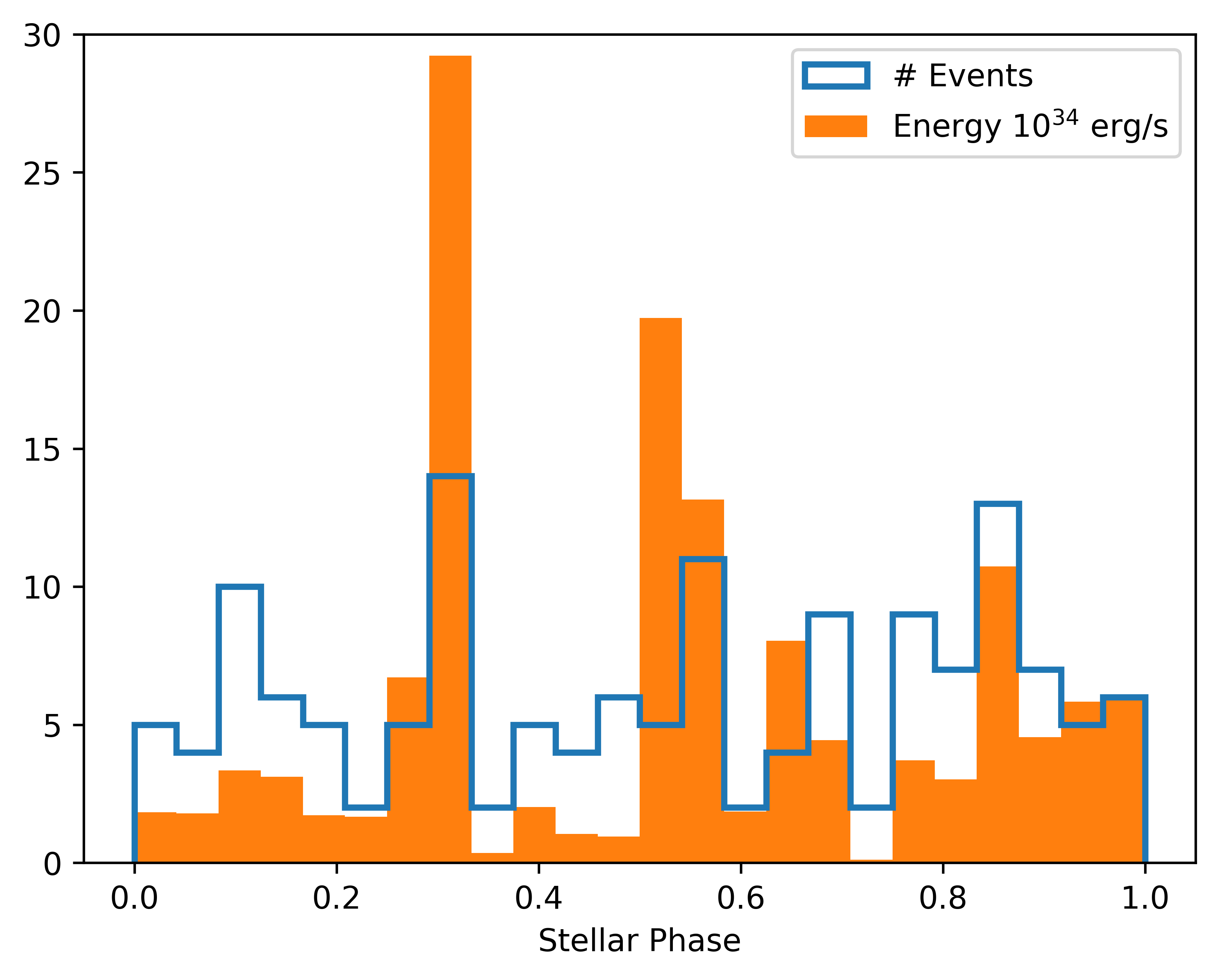}}
    \subfigure[]{\includegraphics[scale =0.5]{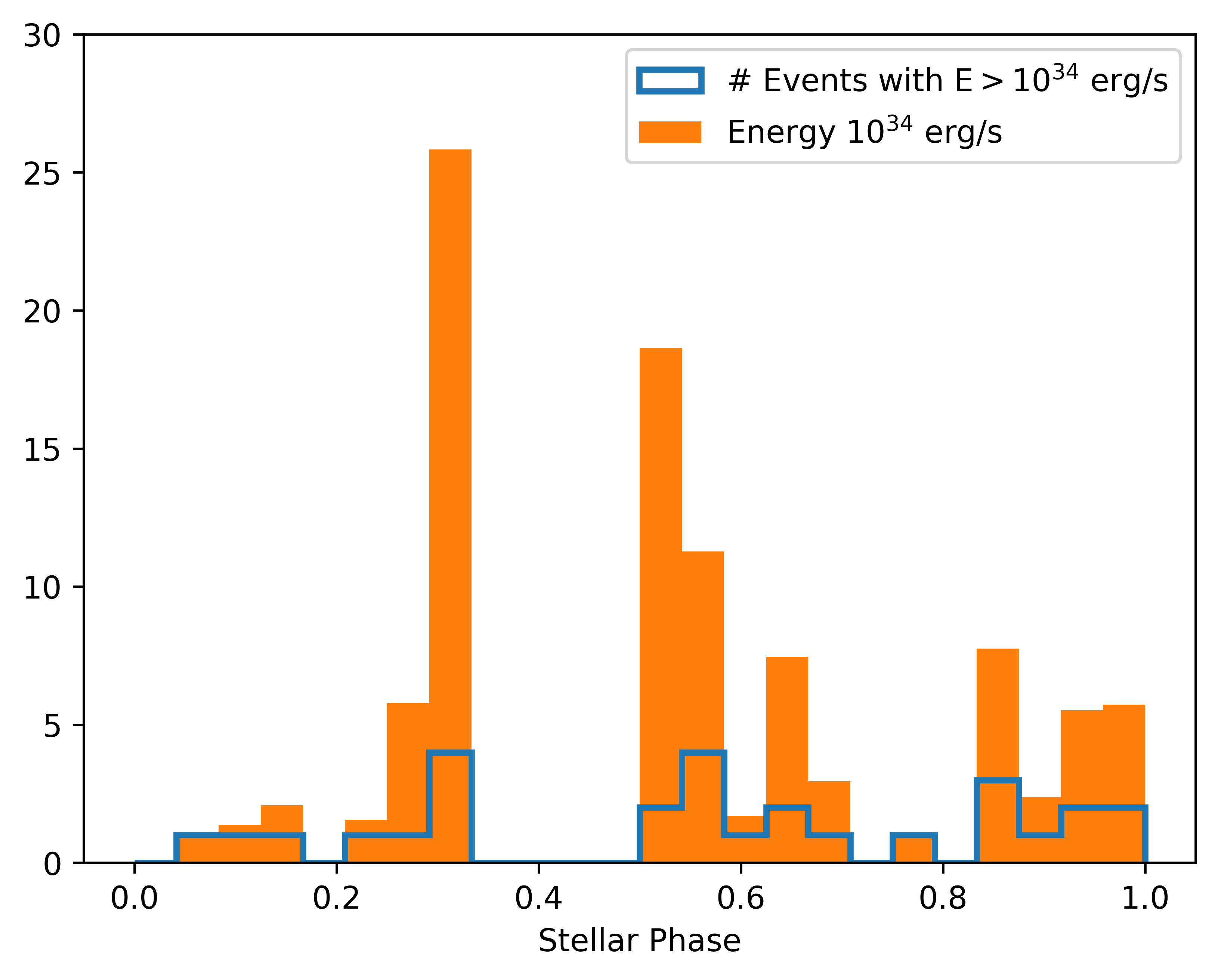}} \\
    \subfigure[]{\includegraphics[scale =0.5]{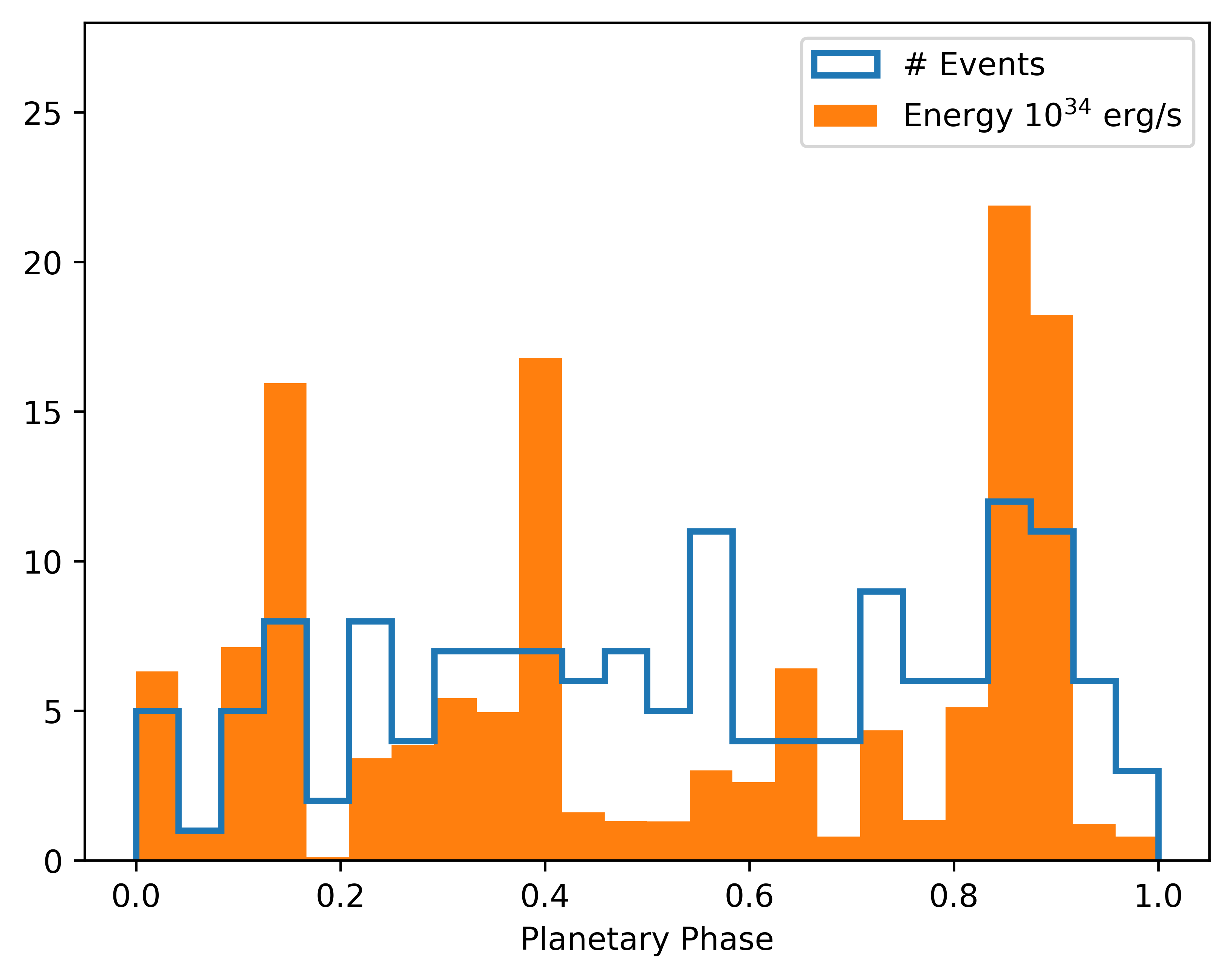}}
    \subfigure[]{\includegraphics[scale =0.5]{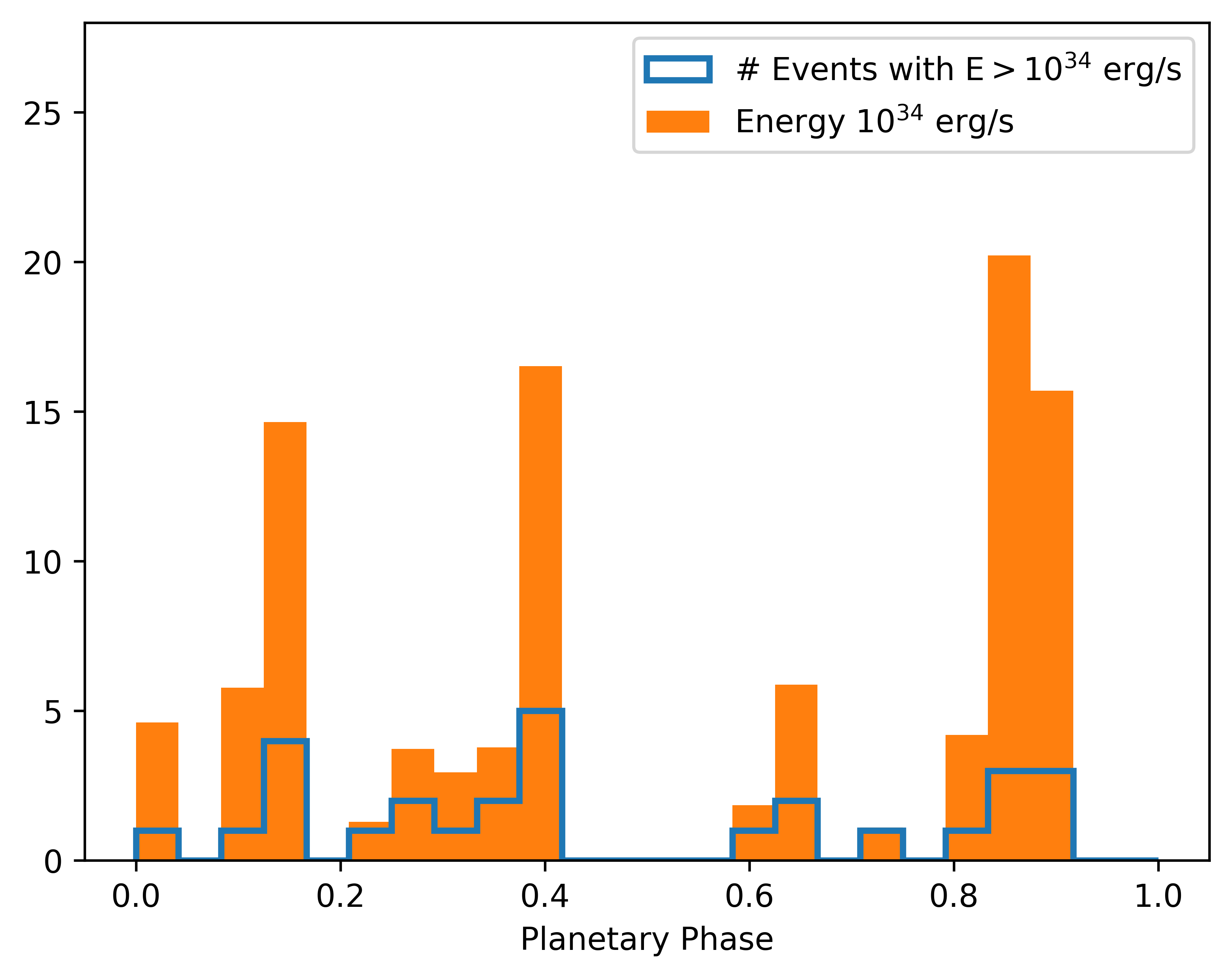}} \\
    
    \caption{Top: left panel (a) histogram for event energy (in orange) and number of events (in blue) per stellar phase for DS Tuc A. Top right panel (b) same as the left panel but for events with energy $E>10^{34}$erg/s. Bottom panels as top panel but for planetary phase.}
    \label{fig:hist_Ds_Tuc}
\end{figure*}

We investigate if short term events are enhanced in particular phases of the stellar rotation cycles or planetary period. Fig. \ref{fig:folding} shows the peaks of each energetic events in the light curve for all the sectors available, phase folded for both stellar and planetary periods (DS Tuc A b and AU Mic b). In all plots, phase = 0 is calculated at the first planetary transit.

\begin{figure*}[!h]
   \centering
   \subfigure[]{\includegraphics[scale =0.5]{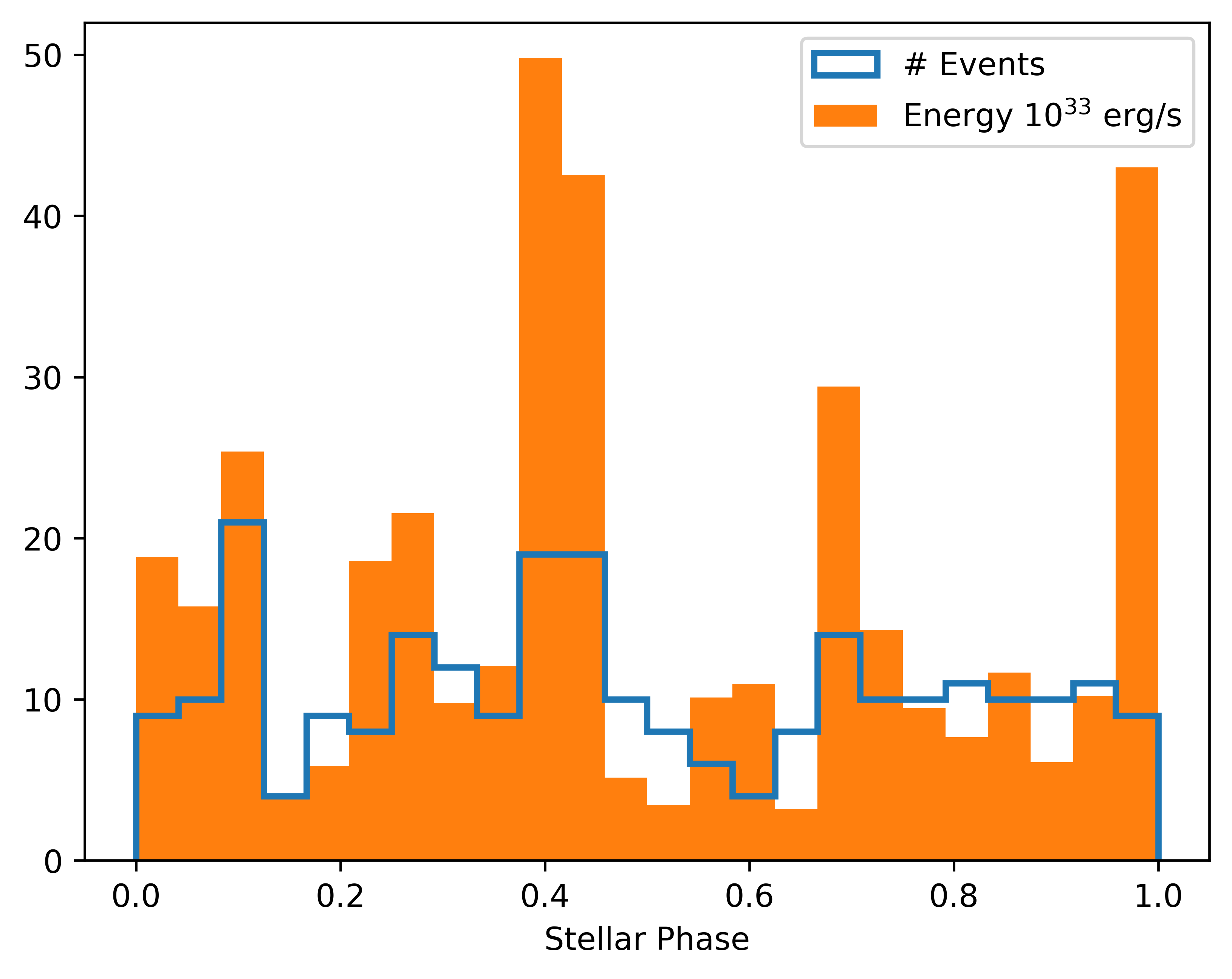}}
   \subfigure[]{\includegraphics[scale =0.5]{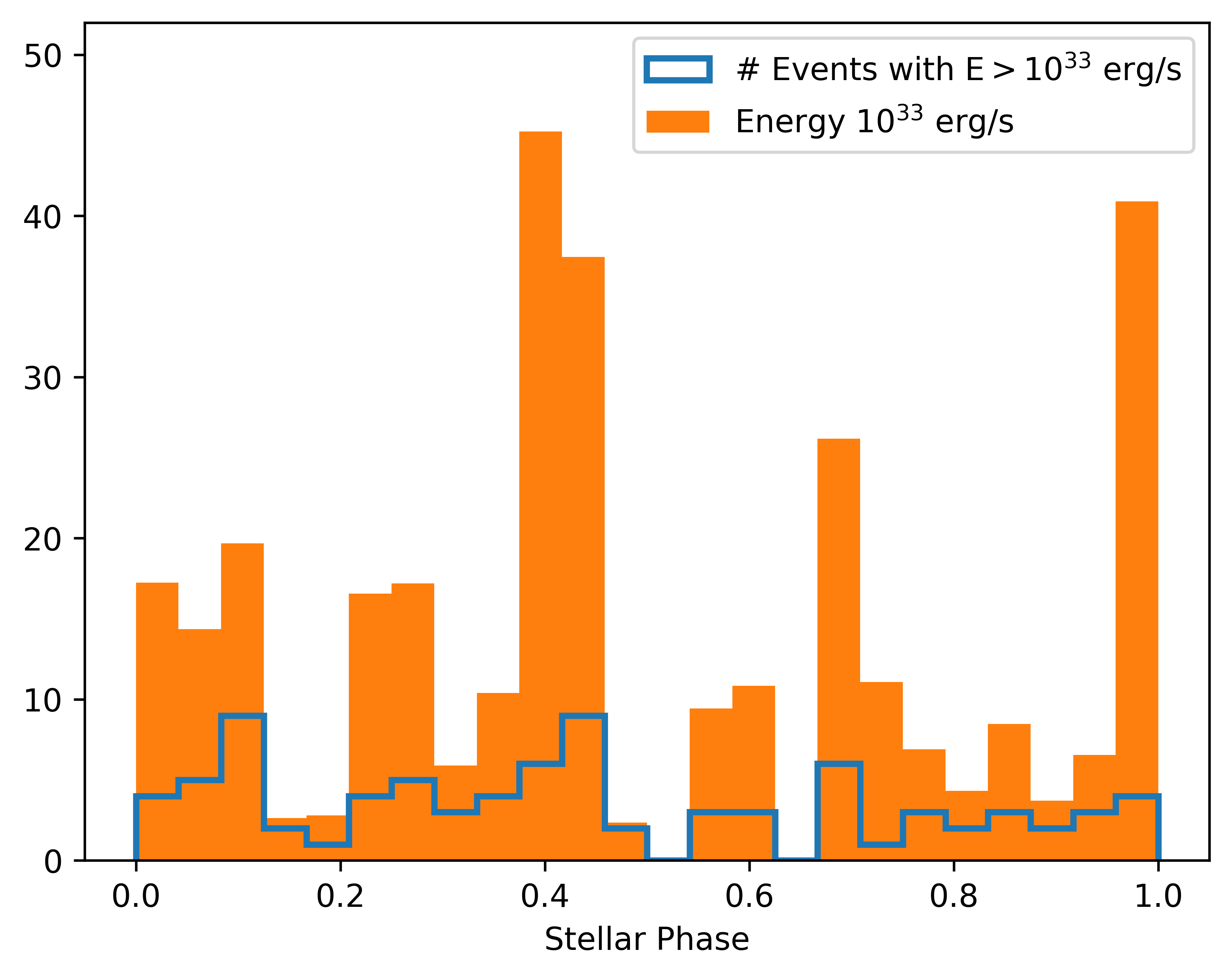}} \\
   \subfigure[]{\includegraphics[scale =0.5]{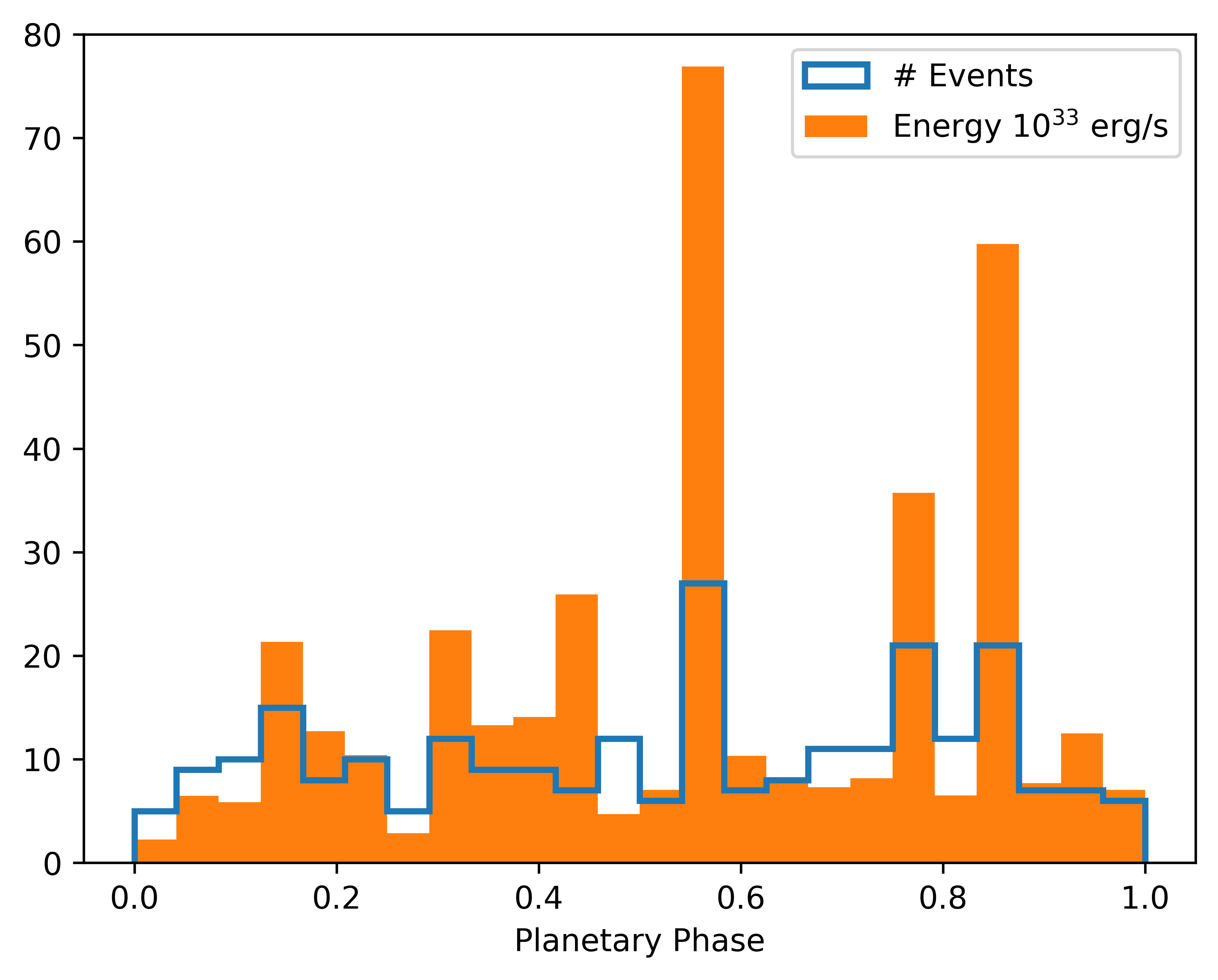}}
   \subfigure[]{\includegraphics[scale =0.5]{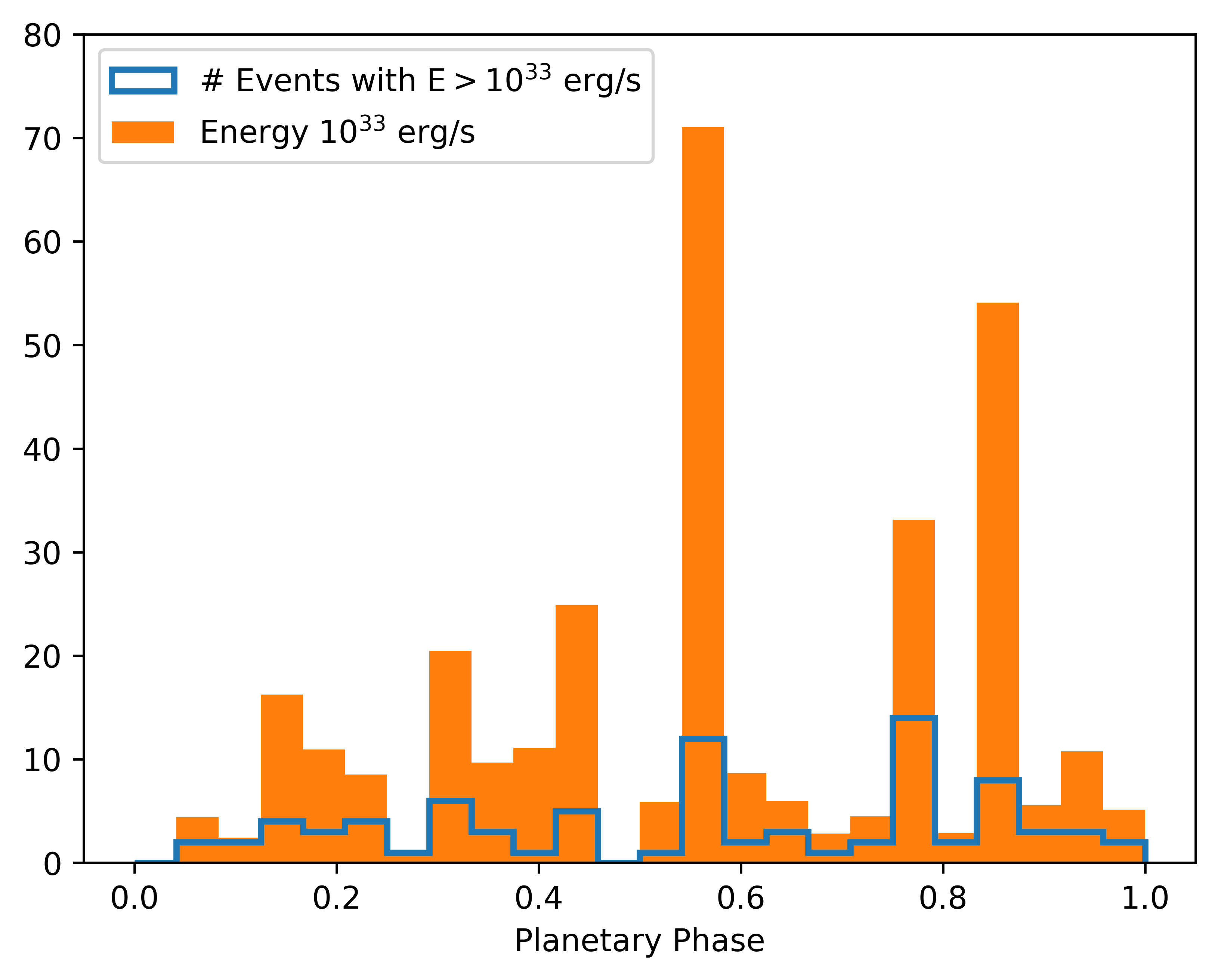}} \\   
   \caption{As Fig.\ref{fig:hist_Ds_Tuc} but for AU Mic.}
    \label{fig:hist_Au_Mic}
\end{figure*}

We also plotted the histograms with the number of events and the total energy per phase bins both for stellar and planetary period. The histograms for both DS Tuc A and AU Mic are showed in Fig. \ref{fig:hist_Ds_Tuc} and Fig. \ref{fig:hist_Au_Mic}.
DS Tuc A shows two peaks in energy emitted, at stellar phases 0.3 and 0.5 (see Fig. \ref{fig:hist_Ds_Tuc} (a, b)). At phase 0.3 at the energy peak corresponds a peak in the number of events (blue line in Fig. \ref{fig:hist_Ds_Tuc}). It is clear from Fig.  \ref{fig:folding} (a) that  3 over 5 the events at phase 0.3 are originated at the same stellar rotation cycle. Moreover, the second peak at phase 0.5 is generated by a single flare as can be seen in Fig. \ref{fig:folding} (a). The data suggest that there is not a favourite region where stellar activity is enhanced on Ds Tuc. 
During the planetary orbit, a bump both in the number of events and in the integrated energy is visible between phase 0.8 and 0.9 (see Fig. \ref{fig:hist_Ds_Tuc} (c, d)). In the plot in Fig.\ref{fig:folding} (b) is clear that the bump in energy is due to only two events, that occurs at different planetary orbits. We conclude that there are no significant evidence of activity in phase with planetary period on DS Tuc A.

Analogous analysis has been made for AU Mic. In Fig.\ref{fig:hist_Au_Mic} (a, b) two peaks at phase 0.4 and 1 are present. The one at phase 1 is the result of a single energetic event and for this reason is not considered in the analysis. The other peak at phase 0.4 is the results of 4 major events occurring at different stellar rotations each of them followed by a series of less energetic events in Sector 27 (See Fig.\ref{fig:folding} c). We claim that in AU Mic there are hints of activity favoured at specific locations on the stellar surface. The behaviour of the activity in phase with the star rotation cycle may be a consequence of the fact that the distribution of the surface magnetic field in young stars is relatively stable for the time scale of the TESS observations \citep[e.g.][]{Folsom2013,Klein2021}. These stars exhibit quite intense magnetic fields whose dynamo processes are not fully understood \citep[e.g.][]{Donati2014,Yu2019,Hill2019}. As a result, there are regions on the stellar surface where magnetic reconnection phenomena are favoured and generate energetic events.

In histograms in Fig. \ref{fig:hist_Au_Mic} (c, d) 2 major peaks in energy are present: at phase $\approx 0.6$ and $\approx 0.8$. In Fig. \ref{fig:folding} (d) is clear that the peak at $\approx 0.8$ is due to a series of events all generated at the same planetary period (same symbol and same color in Fig. \ref{fig:folding} d). However, the peak at $\approx 0.6$ is generated by events from 4 different planetary periods (from a total of 6 full planetary periods observed.). We can conclude that for AU Mic we observe hints of SPI. This is not the first case of flare periodicity connected with the planetary orbit observed by TESS \citep{Howard2021}. Despite this and the numerous theoretical models that describe SPI \citep[e.g.][]{Lanza2018,Lanza2021,Lanza2022}, the case of AU Mic remains unexplained.

In fact, AU Mic b orbits at relatively great distance from the star, then we do not expect a strong interaction between stellar and planetary magnetic field \citep[as the kind described by][]{Lanza2018}. The planet could also interact with the star through the gravitational potential. Tides generated by the planet could perturb the stellar magnetic internal configuration. The perturbations on the magnetic field could enhance the activity at some specific planetary orbital phases \citep{Lanza2022}.
This scenario however, requires a larger planetary mass and closer orbit than the AU Mic b ones.

\begin{figure}
    \centering
     \subfigure[]{\includegraphics[scale=0.5]{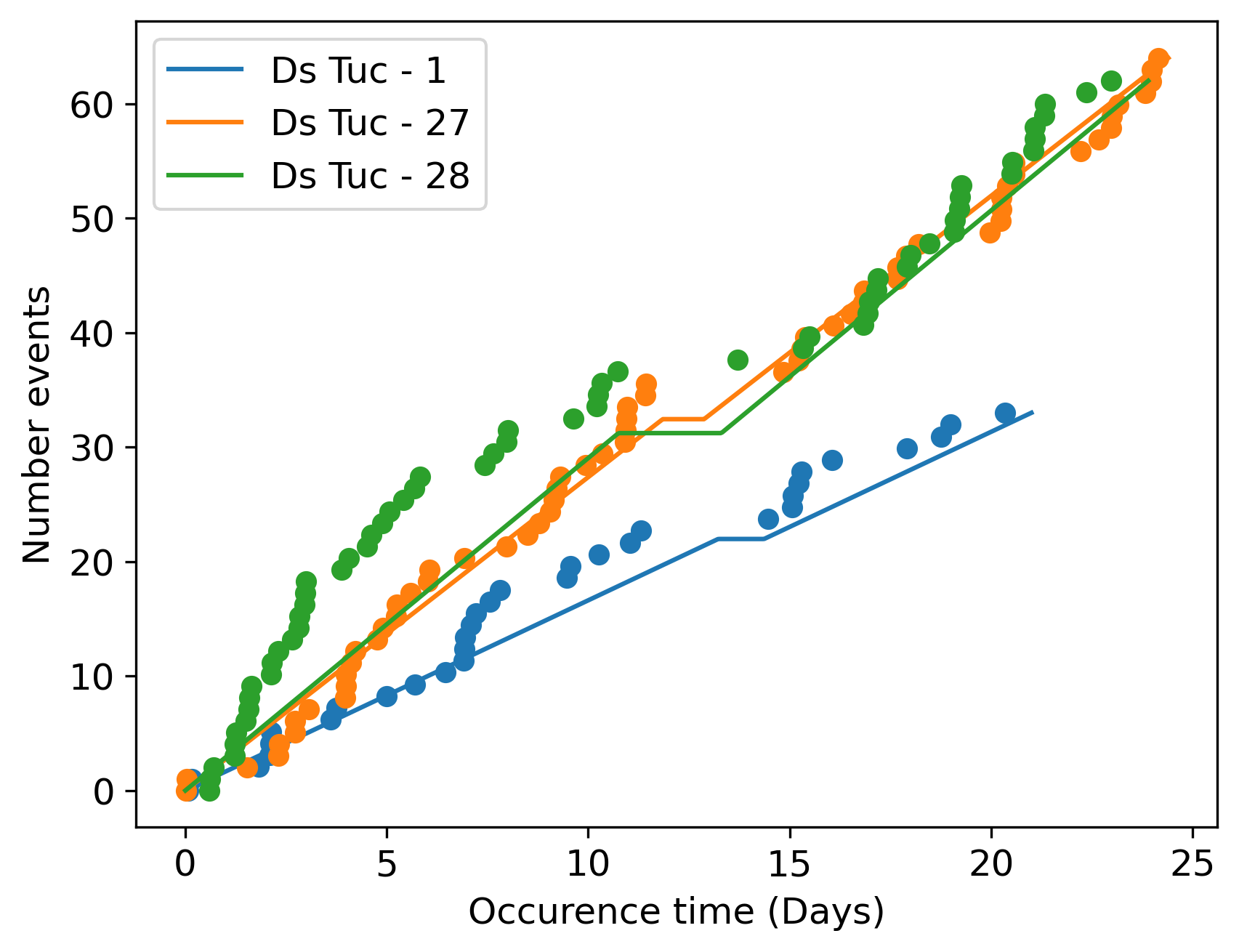}}
     \subfigure[]{\includegraphics[scale=0.5]{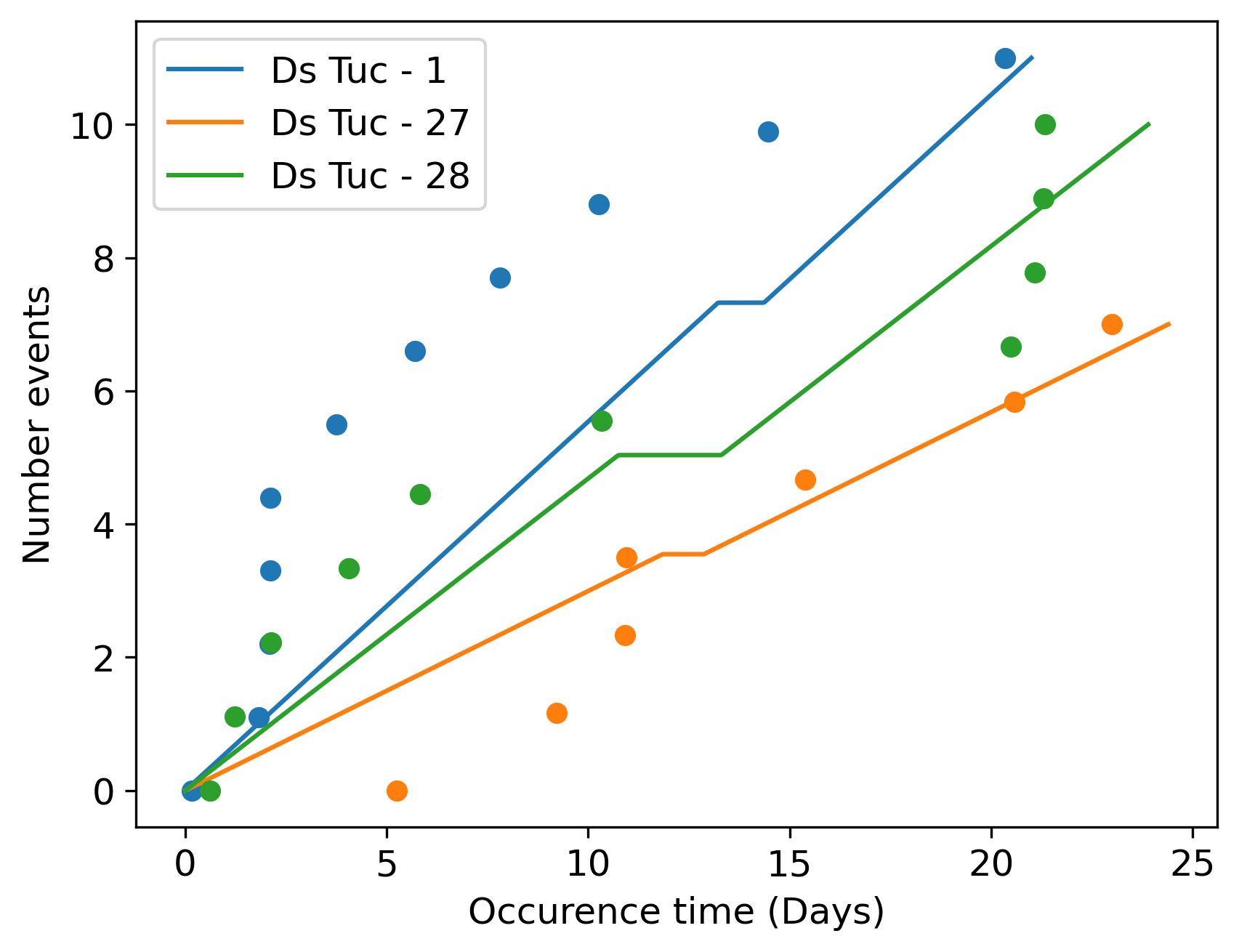}}
    \caption{Expected Vs Observed event distribution during the observed period for the 3 sectors of DS Tuc. The x-axis shows the occurrence time of each energetic event, the y-axis shows the number of the events. The dots represent the real data, the solid lines represent an equally time-spaced distribution. The top panel (a) shows all the events, the bottom panel (b) shows only events with associated energy greater than $10^{34}$erg.  }
    
    \label{fig:exp_vs_obs}
\end{figure}

Energetic events are not evenly distributed during the observation as can be observed in Fig. \ref{fig:exp_vs_obs}. The plot in Fig. \ref{fig:exp_vs_obs} (a) shows the distribution of the energetic events during time. In particular, the events are likely to occur in groups, since the deviation from the uniform distribution. 
To explain this distribution, we envisage two possible explanations: a) it could be a result of the plasma sloshing in a pulse-heated loop \citep{Reale2016}, each wave in the stellar loop generates an increase in the optical light curve and it is identified as energetic events (See Fig. \ref{fig:flare}), or b) a large flare can trigger the occurrence of following new flares.

Fig. \ref{fig:exp_vs_obs} (b) shows the same analysis but only with events with associated energy greater than $10^{34}$erg. The distributions in time for the 3 sectors do not follow either the expected distribution for equally-spaced events, suggesting a link between consecutive energetic events,  supporting hypothesis b). 

\section{Discussion }\label{sec:4}
\begin{figure*}[!h]
    \centering
    \subfigure[]{\includegraphics[scale = 0.5]{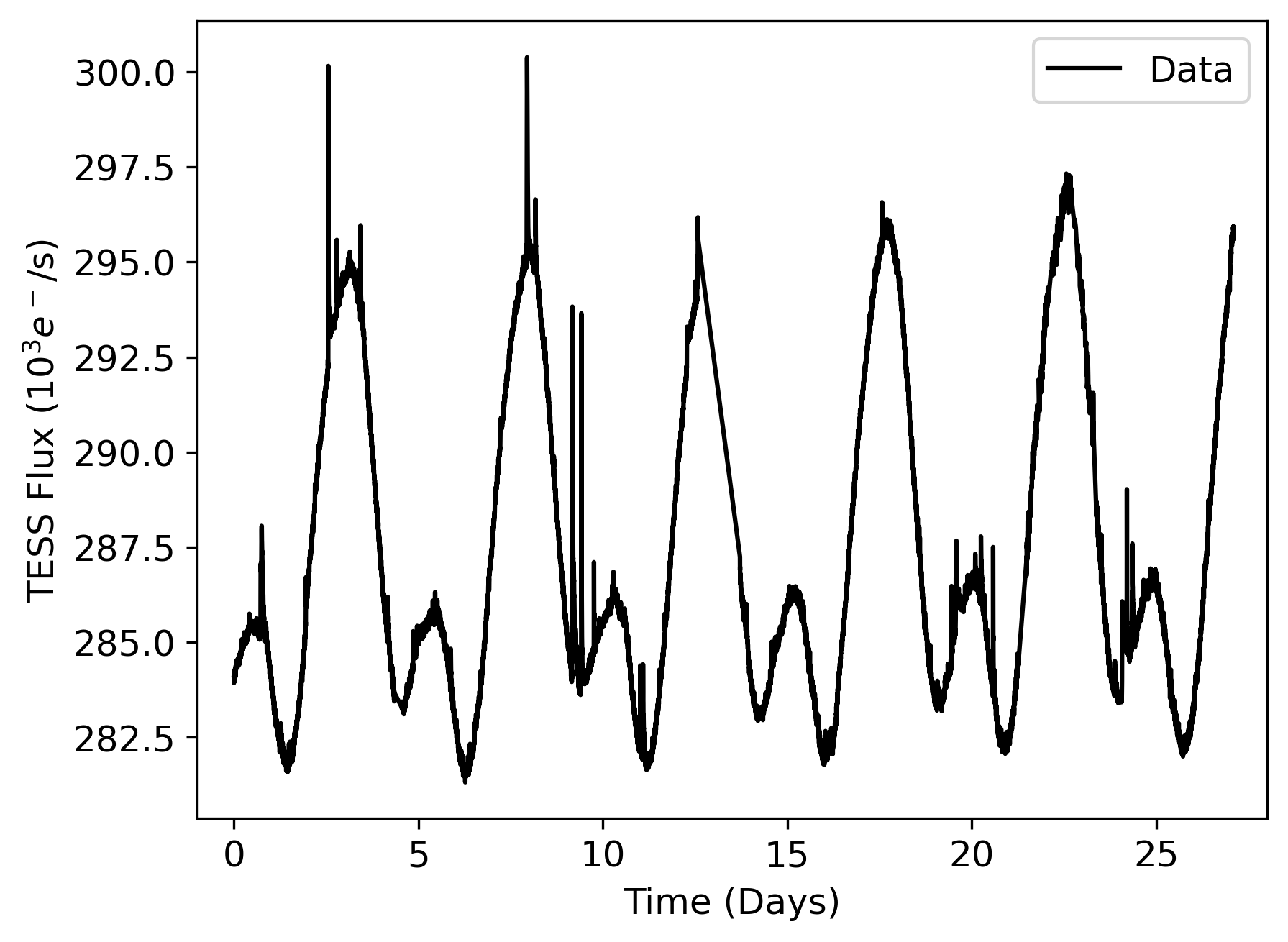}} \subfigure[]{\includegraphics[scale = 0.5]{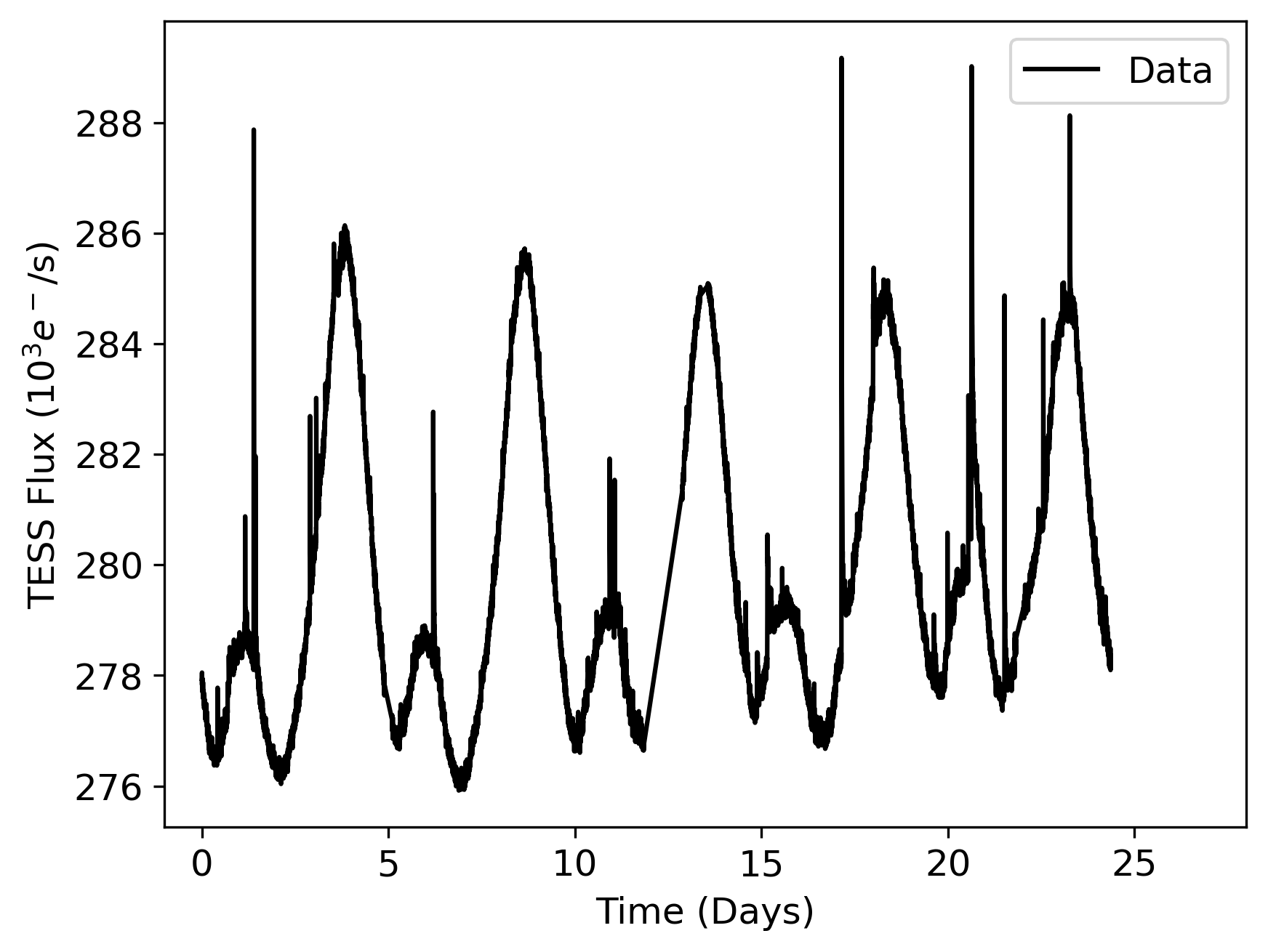}} \\
    \centering
    \subfigure[]{\includegraphics[scale = 0.5]{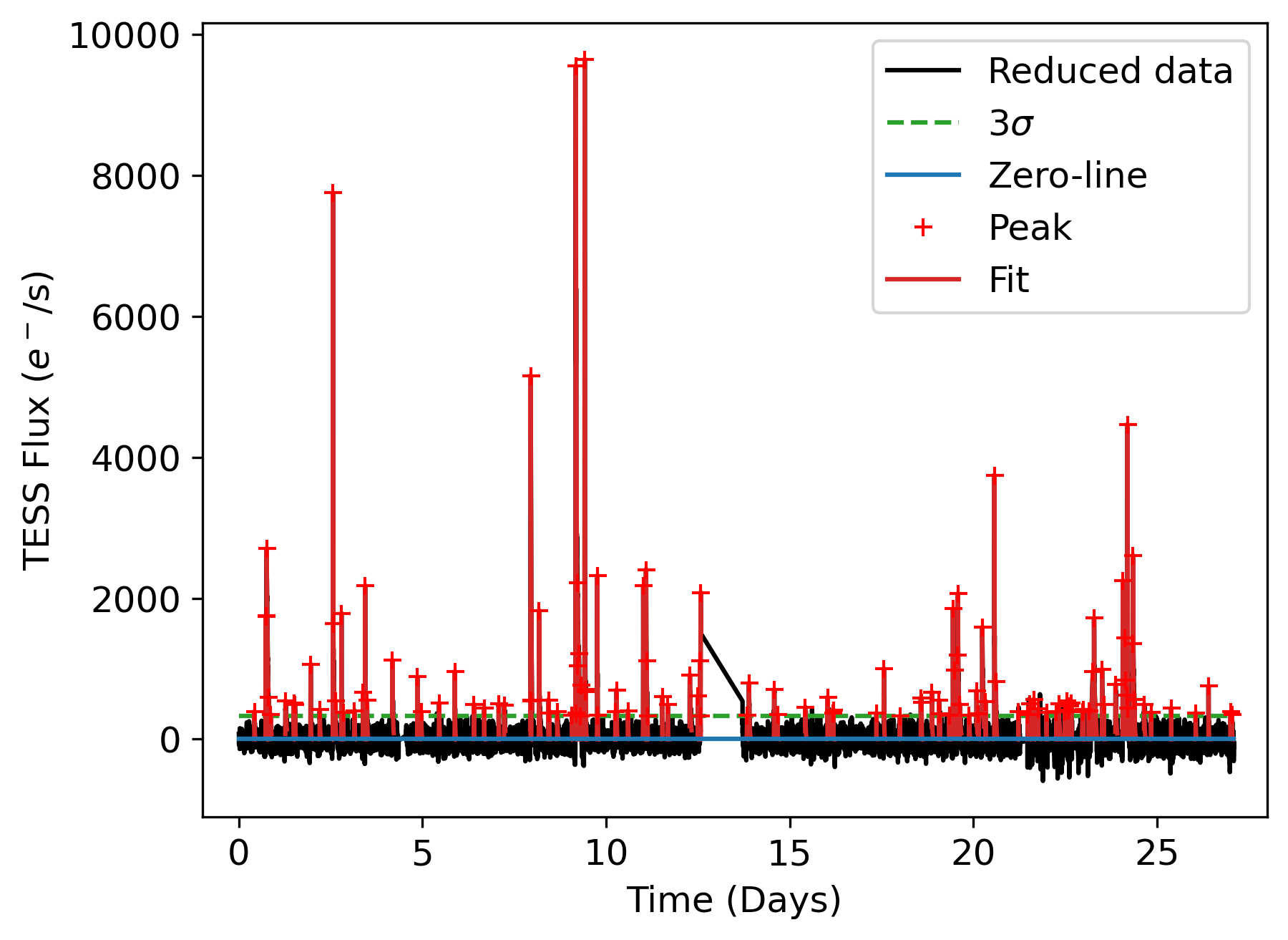}} \subfigure[]{\includegraphics[scale = 0.5]{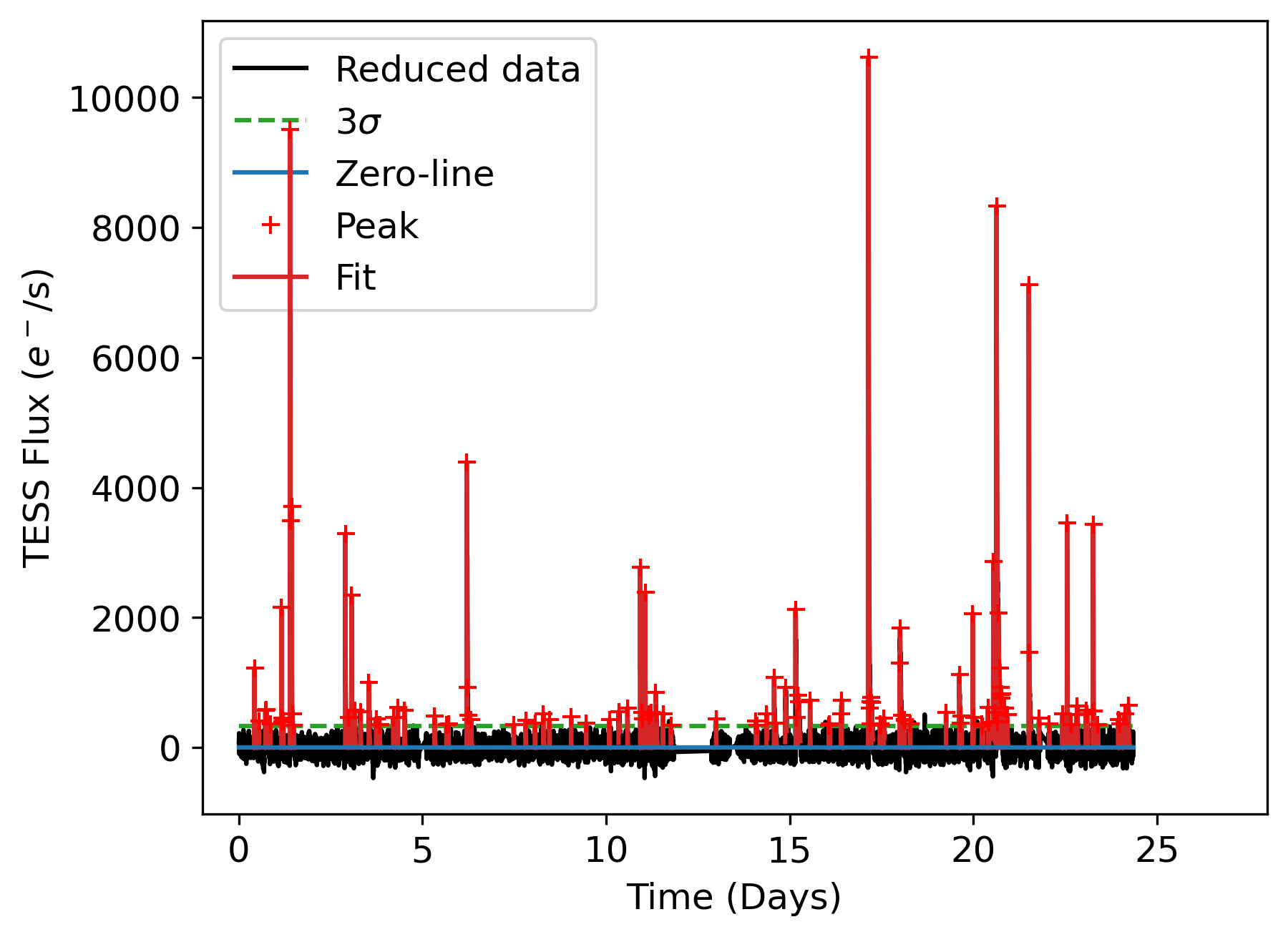}}

    \caption{First step of the analysis of AU Mic. On the top panels (a,b) are the three initial light curves (Sector 1 and 27 respectively). On the bottom panel (c,d) are the final RCs after the GP analysis. The red crosses show the peaks of each event identified and the red lines the fits for each event. The dotted green line shows the $3\sigma$ value and the blue line is the zero-line. }
    \label{fig:lc_gp_aumic}
\end{figure*}

In order to validate our procedure we, also, analyse the  two TESS observations (sectors 1 and 27) of the system AU Mic. We use AU Mic as a test to verify the robustness of our analysis and we compared our results with those present in the literature.
We applied the procedure to AU Mic as described for DS Tuc A and the results are reported in Tab. \ref{tab:last_gp}, while the light curves and RCs are in Fig. \ref{fig:lc_gp_aumic}.

In order to derive the energetic of the identified events, we applied the Eq. {\ref{eq:energy} } adopting a value of $L_{bol} = 0.09 L_{\odot}$ \citep{Plavchan2009}.
According to our analysis, AU Mic shows $\approx 5$ energetic events per day with energy $E>5\times10^{31}$erg.

\cite{Martioli2021} performed an analogous analysis for AU Mic and claim a frequency of 6.35 flares per day, which is comparable to our results. We compared, also, with the work of \cite{Gilbert2021}, they used a technique based on Bayesian inference to detect flares in the AU Mic TESS light curves. Their cumulative curve \citep[Fig. 5 in][]{Gilbert2021} is comparable with those presented herein Fig. \ref{fig:rate} for high energies. However, our technique shows more low energy events than those observed in \cite{Gilbert2021} by a factor $\approx$2. 

AU Mic shows, on average, less energetic events than DS Tuc.
However, is clear from the scatter plot in Fig.\ref{fig:timevsampl} that we are sensitive to lower amplitude (and therefore lower energy) events in AU Mic. This is likely due to the different bolometric magnitude of the two stars. Such difference is evident also in the cumulative curves in Fig. \ref{fig:rate} where the curves appear to be shifted of about one order of magnitude with good sensitivity to events with energy $E > 10^{33}$ erg for AU Mic and $E>10^{34}$ erg for DS Tuc. As discussed in Sect. \ref{sec:3} this depends on the threshold used for defining the peaks of the energetic events, which is affected by the stellar background level when we derive the absolute energy and not relative variations. On the other side, the slope of the flare frequency of the two stars is similar.This implies that even in different stellar spectral types the physics of energetic phenomena is comparable.

The photosphere of DS Tuc A is also more variable than the one of AU Mic, in fact as shown in Tab. \ref{tab:last_gp} the lifetime of active regions for DS Tuc A ($\tau$) is comparable to the rotation period. Thus, the stellar activity level is different for each period. On the contrary, AU Mic shows a lifetime of active regions which is roughly twice the period, this is also evident comparing the light curves in Fig. \ref{fig:lc_gp_aumic} and Fig. \ref{fig:lc_gp}. In fact, DS Tuc A light curves are less regular than those of AU Mic.

\section{Conclusions}\label{sec:5}
Our procedure is able to identify and characterise the short-term variability in light curves outlining the most important physical parameters of each short-term event. Moreover, after the short-term analysis, the procedure can characterise also the long-term stellar variability. We validate the procedure by comparing the results obtained with other techniques by \cite{Martioli2021} for AU Mic, finding a rate of events of $\approx 5$ per day with energy $E_f>5\times 10^{31}$(see Fig. \ref{fig:cumulative}).

In addition to that, we study the system DS Tuc A and find a rate of energetic events of $\approx 2$ events per day with energy greater than $2\times10^{32}$erg/s (see Fig. \ref{fig:rate}). 
The phase analysis of the events suggests that on DS Tuc A there are not preferred stellar or planetary phase for the events (see Fig. \ref{fig:folding}). However, the time distance between the events tend not to be constant, but on the contrary, the events arise in groups (see Fig. \ref{fig:exp_vs_obs}).

The phase analysis of AU Mic shows that may be a favourite region on stellar surface where energetic events are originated. In addition, there are some hints of SPI difficult to explain by current theoretical models.

The results of our analysis could be used as input for precise models of planetary atmosphere chemical evolution and, moreover could be used as a tracer for X-ray flaring activity. In fact, the analysis of XMM-Newton observations that showed at least two relatively large, twin X-ray flares, in a total exposure time of $\sim 70$ks, with clear counterparts in the optical monitor \citep[][Pillitteri et al. in prep]{Benatti2021} suggests that optical flare can be used as a proxy for X-ray flares. 
Assuming that optical flares are a proxy for their X-ray counterparts, from the optical flare distribution we can estimate the X-ray flare distribution using the scaling law between optical and X-ray flare energy from Fig. 3 in \citep{Flaccomio2018}. In particular for X-ray, from Fig. \ref{fig:cumulative}, we expect a frequency of 2 X-ray flares/day with energy $E_x \approx 2\times 10^{31}$ erg.

\begin{acknowledgements}
We thank the anonymous referee for useful suggestions that
have allowed us to improve the paper.
We thank A. F. Lanza for fruitful discussion on SPI models.
We acknowledge financial contribution from the agreement ASI-INAF n.2018-16-HH.0 (THE StellaR PAth project). We acknowledge support from ASI-INAF agreement  2021-5-HH.0 Partecipazione alla fase B2/C della missione ARIEL (Atmospheric Remote-Sensing Infrared Exoplanet Large-survey). This paper includes data collected by the TESS mission. Funding for the TESS mission is provided by the NASA's Science Mission Directorate.
\end{acknowledgements}


\bibliographystyle{aa}
\bibliography{biblio}

\end{document}